\documentclass[review]{elsarticle}

\usepackage{hyperref}

\usepackage{subcaption} 
\usepackage{bm} 
\usepackage{amsmath} 
\DeclareMathOperator*{\argmax}{argmax} 
\usepackage{subdepth} 
\usepackage{mathtools} 

\journal{ }

\usepackage{etoolbox}
\makeatletter
\patchcmd{\ps@pprintTitle}
  {Preprint submitted to}
  {}
  {}{}
\makeatother

\begin{document}

\begin{frontmatter}

\title{Nonstationarity Analysis of Materials Microstructures via Fisher Score Vectors}

\author{Kungang Zhang} 
\author{Daniel W.\ Apley\corref{mycorrespondingauthor}}
\address{Department of Industrial Engineering and Management Sciences, Northwestern University Evanston, IL 60208, USA}

\cortext[mycorrespondingauthor]{Corresponding author}
\ead{apley@northwestern.edu}

\author{Wei Chen}
\address{Department of Mechanical Engineering, Northwestern University Evanston, IL 60208, USA}



\begin{abstract}
Microstructures are critical to the physical properties of materials. Stochastic microstructures are commonly observed in many kinds of materials (e.g., composite polymers, multiphase alloys, ceramics, etc.) and traditional descriptor-based image analysis of them can be challenging. In this paper, we introduce a powerful and versatile score-based framework for analyzing nonstationarity in stochastic materials microstructures. The framework involves training a parametric supervised learning model to predict a pixel value using neighboring pixels in images of microstructures~(as known as micrographs), and this predictive model provides an implicit characterization of the stochastic nature of the microstructure. The basis for our approach is the Fisher score vector, defined as the gradient of the log-likelihood with respect to the parameters of the predictive model, at each micrograph pixel. A fundamental property of the score vector is that it is zero-mean if the predictive relationship in the vicinity of that pixel remains unchanged, which we equate with the local stochastic nature of the microstructure remaining unchanged. Conversely, if the local stochastic nature changes, then the mean of the score vector generally differs from zero. In light of this, our framework analyzes how the local mean of the score vector varies across one or more image samples to: (1) monitor for nonstationarity by indicating whether new samples are statistically different than reference samples and where they may differ and (2) diagnose nonstationarity by identifying the distinct types of stochastic microstructures that are present over a set of samples and labeling accordingly the corresponding regions of the samples. Unlike feature-based methods, our approach is almost completely general and requires no prior knowledge of the nature of the nonstationarities or the microstructure itself. Using a number of real and simulated micrographs, including polymer composites and multiphase alloys, we demonstrate the power and versatility of the approach. 
\end{abstract}


\end{frontmatter}


\section{Motivation and Introduction}
\label{s_SM:moti_intro}
The physical properties of materials depend strongly on their microstructures. The discovery and design of materials with certain complex microstructures and superior or desirable properties are boosted by advances in fabrication and imaging techniques. In general, microstructures can be nonstationary in the sense that the nature of the microstructure varies across a single image or multiple image samples, due to variation in processing conditions, input materials, environmental conditions, etc. For example, state-of-the-art additive manufacturing (AM) gives unprecedented control of the microscopic phases of composite materials to customize the properties resulting in increasingly sophisticated multiphase microstructures~(\cite{bandyopadhyay2018additive}). After fabricating materials, advanced imaging approaches can be used to non-intrusively and efficiently collect massive amount of $2$D and $3$D microscopic image data, examples of which include transmission electron microscopy (TEM)~(\cite{stenley1993scanning}), scanning tunneling electron microscopy (STEM)~(\cite{stroscio1993scanning}), synchrotron-based tomography~(\cite{kinney1992x}), magnetic resonance imaging~(\cite{howle1993visualization}), and confocal microscopy~(\cite{fredrich1995imaging}). Recent literature shows that such advances have led to fast-growing image databases of complex microstructures of materials (also known as micrographs)~(\cite{gagliardi2015material,takahashi2016materials,blaiszik2016materials,puchala2016materials,jain2016new,jain2016research,kim2016organized,rose2017aflux,AFRL2018}), which in turn incentivizes further effort to design general, automated, and efficient workflows to process the micrographs.

\begin{figure}[!htbp]
\centering
\begin{subfigure}[t]{0.2655\linewidth}
         \centering
         \includegraphics[width=\textwidth, trim=.0in .0in .0in .0in, clip]{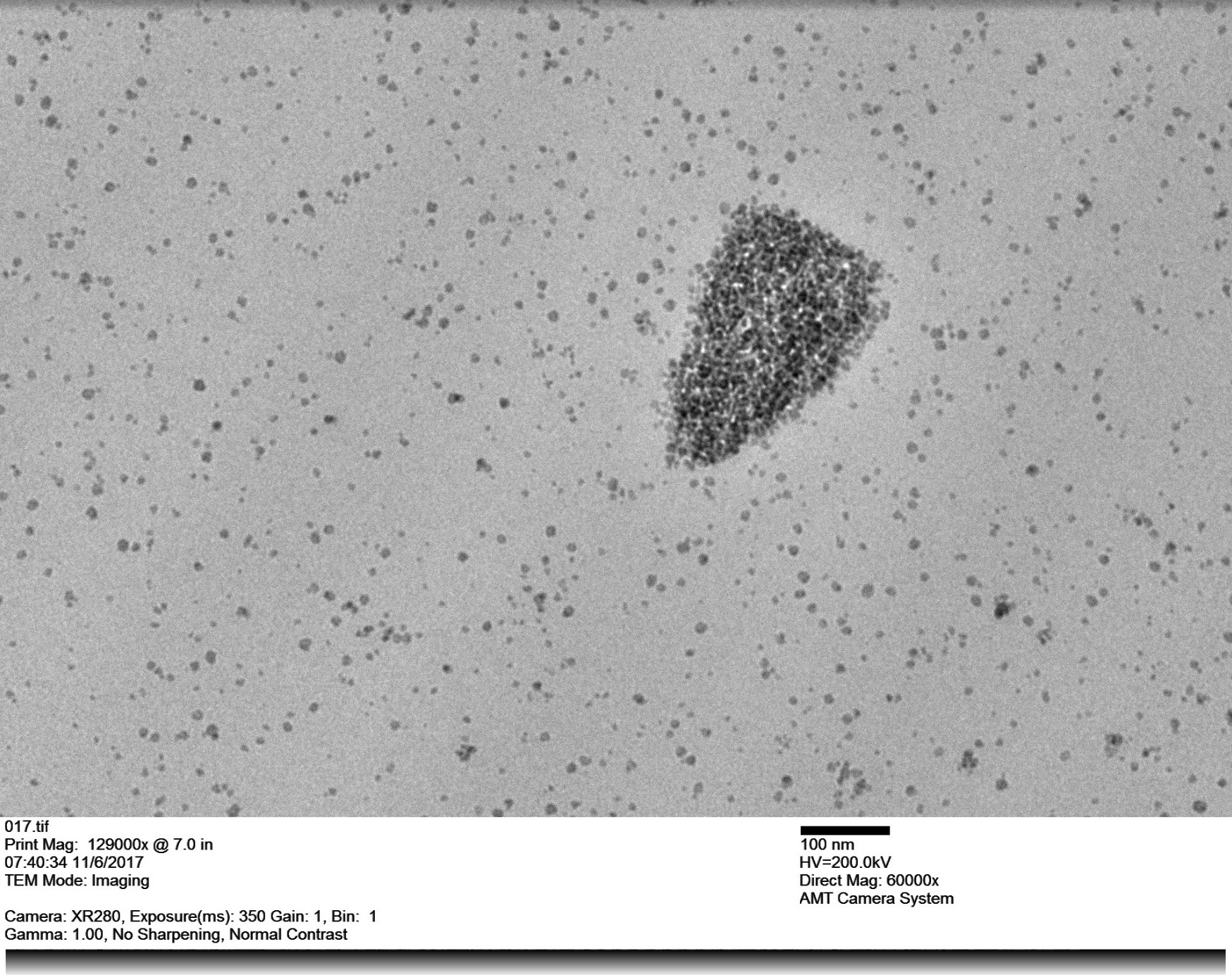}
         \captionsetup{width=.95\linewidth}
         \caption{}
         \label{fig_SM:PMMA}
  \end{subfigure}
  \begin{subfigure}[t]{0.28\linewidth}
         \centering
         \includegraphics[width=\textwidth, trim=.0in .0in .0in .0in, clip]{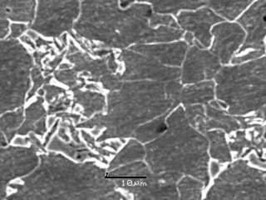}
         \captionsetup{width=.95\linewidth}
	 \caption{}
         \label{fig_SM:dual_phase_steel}
  \end{subfigure}
  \begin{subfigure}[t]{0.21\linewidth}
         \centering
         \includegraphics[width=\textwidth, trim=.0in .0in .0in .0in, clip]{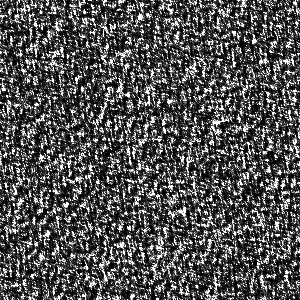}
         \captionsetup{width=.95\linewidth}
	\caption{}
         \label{fig_SM:ar_model1}
  \end{subfigure}
  \begin{subfigure}[t]{0.21\linewidth}
         \centering
         \includegraphics[width=\textwidth, trim=.0in .0in .0in .0in, clip]{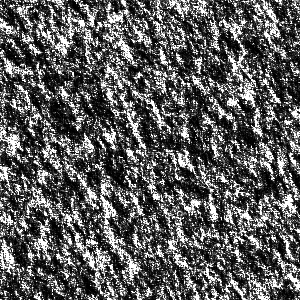}
         \captionsetup{width=.95\linewidth}
	\caption{}
         \label{fig_SM:ar_model2}
  \end{subfigure}
\caption{Examples of micrographs of microstructures. (a) A TEM image of silica particles in Polymethyl methacrylate (PMMA). (b) A SEM image of dual-phase steel~(\cite{banerjee2013segmentation}). (c)-(d) Two images of simulated microstructures.}
  \label{fig_SM:mat_example1}
  \end{figure}

An important problem for materials scientists and manufacturing/processing engineers is to automatically analyze nonstationarity of microstructure image samples for quality control purposes (e.g., to detect instability in the material processing that inadvertently results in changes in the microstructure and material properties) and, more generally, to more fully understand the nature of the material being produced~(\cite{bui2018monitoringa,bui2018monitoringb,liu2006estimation,lin2007computer,bharati2003softwood}). By ``nonstationarity", we mean that the stochastic nature of the microstructure varies spatially, either within a single image sample or across multiple image samples. For example, Figure~\ref{fig_SM:dual_phase_steel} is an SEM image of steel showing two distinct phases, each of which corresponds to a distinct stochastic nature. And Figures~\ref{fig_SM:ar_model1} and~\ref{fig_SM:ar_model2} show two simulated microstructure samples that, although stationary within each sample, are nonstationary across the two samples. In Sections~\ref{s_SM:model_ms} and~\ref{s_SM:methods}, we more formally define nonstationarity in the context of a predictive supervised learning model fitted to the micrograph.

For microstructures whose behavior depends predominantly on clearly defined geometric features (e.g., particle size, volume fractions, particle distance, or simple inclusion shapes, etc.), one could directly apply a number of standard quality control monitoring methods~(\cite{montgomery2007introduction}) and/or profile monitoring methods~(\cite{paynabar2016change,grasso2016using,viveros2014monitoring}). However, such methods lack generality in the sense that they monitor only a specific set of predefined features and cannot detect more general changes in the microstructure, and they are often not applicable to common microstructures with more stochastic natures~(\cite{torquato2002statistical}), for examples metals~(\cite{lewandowski2016metal}), polymer composites~(\cite{ligon2017polymers}), ceramics~(\cite{chen20193d,moritz2018additive}), etc. Such stochastic microstructures do not have well-defined features that provide a complete characterization of the microstructure, as illustrated in Figure~\ref{fig_SM:mat_example1}. 

Previous approaches to analyzing these types of stochastic microstructures, such as $N$-point correlation functions, lineal path functions, spectral density through Fourier transformation, and joint probability functions~(\cite{efros1999texture,lu1992lineal,oren2003reconstruction,torquato2006necessary,jiao2008modeling,liu2013computational}) use certain statistical features. These methods provide a richer representation of the stochastic nature than geometric and descriptor-based features, but have serious limitations~(\cite{cang2018improving}). First, the statistical features are only summary statistics for the full joint distribution of the stochastic microstructure and do not provide a sufficiently complete representation for some complex microstructures. For example, a $2$-point correlation function is an incomplete statistical description and can also be computationally prohibitive for large images~(\cite{decost2015computer}), and multi-point correlation functions are even more computationally prohibitive while still not providing a complete characterization. In general, there is a fairly severe tradeoff that must be balanced between generality and completeness of the statistical representation on the one hand, versus tractability and computational feasibility achieved by working with only a subset of the statistical features on the other hand; and balancing this tradeoff is complicated by the fact that the most appropriate subset of features is always material-dependent. Second, the most common statistical features are global features in the sense that they are computed over an entire image sample (or at least a sufficiently large subregion of the sample) and therefore cannot capture local behavior that varies on a finer scale. 

In contrast, the score-based framework that we introduce in this paper is almost completely general and, in theory, entails a complete statistical representation of even the most complex stochastic microstructures. It also provides a pixel-by-pixel measure of nonstationarity and can monitor fine-scale local changes in the microstructure. Finally, because the score vectors are automatic by-products of the predictive model training, the computational expense is reasonable. 

One component of our framework uses the approach of~\cite{bostanabad2016stochastic} and \cite{bui2018monitoringa,bui2018monitoringb} to characterize the stochastic nature of microstructures. More specifically, the conditional distribution of each pixel value given its neighboring pixels is approximated by training a supervised learning model to predict the pixel value. In other words, the extremely high-dimensional joint distribution of the stochastic microstructure is implicitly represented by a trained supervised learning model for predicting an individual pixel value given its neighbors. For monitoring for global changes in the stochastic nature from sample to sample, \cite{bui2018monitoringa} trained a separate supervised learning model on each image sample and then used a form of generalized likelihood ratio test to detect changes (i.e., nonstationarity) in the stochastic nature from sample to sample. However, their approach cannot monitor or diagnose nonstationarity within a sample, in addition to being computationally expensive due to the need to fit a separate supervised learning model for each sample. For detecting local anomalies or defects, \cite{bui2018monitoringb} fitted a single supervised learning model to a reference image sample. They then computed various statistics on the residuals of the predictive model (each pixel has a predicted value and a residual error) to detect local defects like voids, tears, etc. However, their approach cannot monitor or diagnose general nonstationarity beyond the presence of certain types of local defects. 

In this study, we develop a powerful and versatile framework based on Fisher score vector concepts to analyze general nonstationarity within and across samples. The Fisher score vector is defined as the gradient of the log-likelihood with respect to the parameters of the fitted predictive model, and each pixel is associated with its own score vector. In Section~\ref{s_SM:methods}, we discuss a fundamental property of the score vectors that provides the basis for our approach. Namely, under fairly general conditions, the score vector is zero-mean if the predictive relationship in the vicinity of the pixel remains unchanged from the training sample on which the predictive model was fitted, and it is nonzero-mean otherwise. Because we equate changes in the stochastic nature of the microstructure with changes in the predictive model that implicitly represents the stochastic nature, our framework for analyzing nonstationarity reduces to analyzing the local mean of the score vector as it varies spatially across one or more image samples. We apply these concepts to two different but related contexts in Sections~\ref{s_SM:supervised_moni} and~\ref{s_SM:unsupervised_chara}. In Section~\ref{s_SM:supervised_moni}, we develop a nonstationarity monitoring (NM) method, the purpose of which is to indicate whether new samples are stationary and statistically equivalent to some reference samples and, if they are not, at which spatial locations they may differ. In Section~\ref{s_SM:unsupervised_chara}, we develop a nonstationarity diagnostic (ND) method, the purpose of which is to identify the distinct types of stochastic microstructure behavior present over a single sample or set of samples and label each region of the samples according to the microstructure type. Using real (e.g., Figures~\ref{fig_SM:PMMA} and~\ref{fig_SM:dual_phase_steel}) and simulated (e.g., Figures~\ref{fig_SM:ar_model1} and~\ref{fig_SM:ar_model2}) material examples, we illustrate use of the framework and demonstrate its power and versatility.

\section{Background on Characterizing the Stochastic Nature of Microstructures via Supervised Learning}
\label{s_SM:model_ms}
In this section, we review a general approach for modeling the stochastic nature of microstructures with micrographs of materials~(\cite{bostanabad2016stochastic}). For stochastic microstructures, we can view each microstructure image sample as a realization of some underlying spatial random process. Let $\bm{X}=[X_1, X_2, \cdots, X_m]^T$ denote the concatenated vector of pixel values of a micrograph (or set of micrographs), where $m$ is the number of pixels in the micrograph(s). Let $P (\bm {X})$ denote the joint distribution of $\bm {X}$. If it were available, the joint distribution $P (\bm {X})$ completely characterizes the stochastic nature of the microstructure sample $\bm {X}$, and distinguishing two statistically different samples amounts to distinguishing their joint distributions.

Directly modeling the distribution is of course computationally prohibitive. Two assumptions that reduce the problem to a more tractable one are \textit{the Markov locality property} and \textit{stationarity}~(\cite{efros1999texture,levina2006texture}) over the sample or at least over a subregion of the sample. The Markov locality property assumes that the conditional distribution of the $i^{\mathrm{th}}$ pixel value $X_i$, given some appropriate set of its neighboring pixels $\bm{\mathcal{N}}(X_i)$, does not depend on the remaining pixels in the sample, i.e., that $P_i(X_i|\bm{X}_{-i}) \equiv P_i(X_i|\bm{\mathcal{N}}(X_i))$, where $\bm{X}_{-i}$ denotes the vector $\bm {X}$ but excluding the entry $X_i$. Stationarity means that the conditional distribution is independent of the location of a pixel, i.e., $P_i(X_i=y|\bm{\mathcal{N}}(X_i)=\bm{x}) \equiv P(X_i=y|\bm{\mathcal{N}}(X_i)=\bm{x})$ is the same function of the tuple $(y, \bm {x})$ for all locations $i$ in the region over which the image is stationary. Then, we can think of a stationary sample $\bm {X}$ as being a realization generated by the conditional distribution $P(X|\bm{\mathcal{N}}(X))$ via a mechanism analogous to Gibbs sampling. In other words, the compact predictive model $P(X|\bm{\mathcal{N}}(X))$ implicitly represents the stochastic nature of the micrograph $\bm{X}$ and can be treated as the ``fingerprint" of a microstructure. Modeling this conditional distribution by fitting some supervised learning model to predict $ {X}$ given $\bm{\mathcal{N}}(X)$ is a tractable problem, and if we can effectively learn this model through image sample data, we can detect and analyze changes in the stochastic nature of the microstructure via detecting and analyzing changes in the conditional distribution $P(X|\bm{\mathcal{N}}(X))$. The remainder of the paper develops our approach for accomplishing this. 

\section{Fundamental Theory for Score-Based Nonstationarity Monitoring and Diagnostics}
\label{s_SM:methods}

In this section, we first relate microstructure nonstationarity to stationarity of the parametric supervised learning model that represents $P(X|\bm{\mathcal{N}}(X))$ in Section~\ref{ss_SM:param_super_model}. Then, the main concepts behind our score-based framework for analyzing nonstationarity are discussed in Section~\ref{ss_SM:fund_theory_score}. We discuss a spatial smoothing technique for estimating the local mean of the score vectors in Section~\ref{ss_SM:spa_ave_score}. 

\subsection{Representing Microstructure Nonstationarity via a Parametric Supervised Learning Model}
\label{ss_SM:param_super_model}
As discussed in Section~\ref{s_SM:model_ms}, the conditional distribution $P(X|\bm{\mathcal{N}}(X))$ can be treated as a ``fingerprint" of the stochastic nature of the microstructure. To model $P(X|\bm{\mathcal{N}}(X))$, we fit a parametric supervised learning model to the data $\{(X_i,\bm{\mathcal{N}}(X_i))\}_{i=1}^m$ from one or more training micrographs, which we will temporarily treat as stationary. The approximated (learned) conditional distribution is denoted as $P(y|\bm{x};\bm{\theta})$, where $\bm{\theta}$ is the vector of parameters of the supervised learning model (e.g., the set of weights for all nodes in a neural network), $y$ is the value of a target pixel, and $\bm{x}$ are the values of its neighboring pixels. Consider grayscale images, which are a common form of microstructure image data. As a tractable and convenient means to an end, we model the conditional distribution as normal with mean $g (\bm {x};\bm { \theta})$ and variance $ \sigma^2$, where $g(\bm{x};\bm{\theta})$ is the parametric supervised learning model for predicting the mean of the pixel, and $\sigma^2$ is the variance of the residual errors of the predictions. Under certain identifiability assumptions, it is reasonable to treat the parameter vector $\bm{\theta}$ of a given parametric model as the ``fingerprint" of the microstructure. If we define $\bm{\theta}_{i},i\in\{0,1\}$, as the ``true" values of parameters for two micrographs or two regions within one micrograph, nonstationarity (stationarity) across the two regions can be represented as $\bm{\theta}_0 \neq \bm{\theta}_1$ ($\bm{\theta}_0 = \bm{\theta}_1$). There are a number of desirable aspects of this modeling procedure. First, the model provides a concise representation of the microstructure regardless of the size of the training data and can be easily applied to new image samples. Second, it can represent general stochastic microstructures and does not require or involve any microstructure-specific knowledge or features. Third, it provides a well-defined mathematical representation of nonstationarity of stochastic microstructures that is flexible in the sense that it can be used in conjunction with any parametric supervised learning model that suitably models the microstructure. Lastly, the score vectors (defined in Section~\ref{ss_SM:fund_theory_score}) of each observation can be easily computed and are by-products of model training or fine-tuning via the popular stochastic gradient descent (SGD) or related algorithms, which results in our framework having reasonable computational expense. We elaborate on these aspects in subsequent sections.   

In this study, we demonstrate our framework with grayscale microstructure images, for which predicting pixel values is a regression problem. The framework can be easily extended to classification problems which correspond to micrographs for which each pixel has been converted to categorical values that indicate to which phase the pixel belongs. In this classification setting, the supervised learning model directly produces $P(X|\bm { \mathcal{N}}(X))$, which are the multinomial probabilities of the pixel being in each phase. 

\subsection{The Score Vector and Its Zero-Mean Property Under Stationarity}
\label{ss_SM:fund_theory_score}
Our framework for microstructure nonstationarity analysis is inspired by recent work on score-based concept drift monitoring~(\cite{zhang2020concept}) for detecting temporal changes in predictive relationships with data collected over time. For a given model $P(X|\bm{\mathcal{N}}(X))$, the score function/vector associated with an individual pixel $X$ (having value $y$ and neighborhood values $\bm{x}$) is formally defined as the gradient of the log-likelihood:
\begin{align}
\bm {s}(\bm {\theta}; y, \bm{x}) = \nabla_{\bm {\theta}} \log P(X = y | \bm{\mathcal{N}}(X) = \bm{x}; \bm {\theta})
\label{eqn_SM:score_func}
\end{align}
where $P(X = y | \bm{\mathcal{N}}(X) = \bm{x}; \bm {\theta})$ is our parametric conditional likelihood for an individual observation $(y, \bm {x})$ and $\nabla_{\bm{\theta}}$ is the gradient operator with respect to the parameters $\bm { \theta}$. According to a fundamental property of score functions~(Proposition $3.4.4$ from~\cite{bickel2015mathematical}) if certain regularity and identifiability conditions are met and if the parametric likelihood is the correct model with a true parameter vector $\bm{\theta}$, then the expectation of the score function evaluated at the true parameter is zero, i.e.,
\begin{align}
E_{\bm {\theta}} [\bm {s}(\bm {\theta}; X, \bm{\mathcal{N}}(X))| \bm{\mathcal{N}}(X)] \vcentcolon = \int \bm {s}(\bm {\theta}; y, \bm{\mathcal{N}}(X))  P(y|\bm{\mathcal{N}}(X); \bm {\theta}) dy = \bm {0} .
\label{eqn_SM:score_mean_zero}
\end{align}
In other words, assuming that the reference micrographs are realizations of the same stationary spatial random process whose conditional distribution is correctly represented by $P(y|\bm {x}; \bm { \theta})$, the expectation of the score vector for each pixel is zero. Note that the expectation in Equation~(\ref{eqn_SM:score_mean_zero}) is conditioned on a specific set of neighbor pixels $\bm{\mathcal{N}}(X)$, and Equation~(\ref{eqn_SM:score_mean_zero}) holds for any such $\bm{\mathcal{N}}(X)$. Consequently the unconditional expectation is also zero, i.e.,
\begin{align}
E_{\bm {\theta}} [\bm {s}(\bm {\theta}; X, \bm{\mathcal{N}}(X))] = \bm {0}
\label{eqn_SM:score_mean_zero_theory}
\end{align}

In real data sets, the expectation in Equation~(\ref{eqn_SM:score_mean_zero_theory}) is replaced by the empirical mean, which should also be $\bm{0}$:
\begin{align}
\begin{aligned}
&\hat{E}_{\bm { \theta}}[\bm{s}(\hat{\bm{\theta}};X, \bm{\mathcal{N}}(X))] \vcentcolon=\frac{1}{m}\sum_{i=1}^{m}\bm{s}(\hat{\bm{\theta}};y_i, \bm{x}_i)=\bm{0}\mathrm{,~where} \\
&\hat{\bm{\theta}} \vcentcolon =  \argmax_{\bm{\theta}}\hat{E}_{\bm { \theta}}[\log{P(X|\bm{\mathcal{N}}(X); \bm{\theta})}] \vcentcolon= \argmax_{\bm{\theta}}\frac{1}{m}\sum_{i=1}^m \log{P(X=y_i|\bm{\mathcal{N}}(X)=\bm{x}_i;\bm{\theta})},
\end{aligned}
\label{eqn_SM:score_exp_zero_emp}
\end{align}
and the operator $\hat{E}_{\bm { \theta}}$ denotes a sample average over the training data $\{(y_i, \bm{x}_i)\}_{i=1}^m$ (here, $y_i$ is the observed value of the $i^{\mathrm{th}}$ pixel, and $\bm{x}_i$ is the vector of observed values of all pixels in the neighborhood of the $i^{\mathrm{th}}$ pixel, and $m$ is the number of pixels in the reference training micrographs that are enough far away from the image boundary to have full neighborhoods) generated under the true parameter vector $\bm{\theta}$, and $\hat{\bm{\theta}}$ is the maximum-likelihood estimator (MLE) of $\bm{\theta}$ for the training data.

Equation~(\ref{eqn_SM:score_exp_zero_emp}) generally holds for training micrographs no matter whether the parametric conditional distribution (or model) is correct. That is because the estimated parameter vector $\hat {\bm { \theta}}$ is the optimum solution when the empirical mean of the log-likelihood is maximized, the gradient of which is the empirical mean of the score vectors at $\hat{\bm{\theta}}$, which will be the zero vector since $\hat{\bm{\theta}}$ is the maximizer. When micrographs or micrograph regions are statistically different from the training/reference micrographs, the parameters $\hat {\bm { \theta}}$ estimated for the reference micrograph generally no longer provide the best fit for the statistically different regions. In such regions the score vector, which is also collinear to the gradient vector of the log-likelihood for individual observations, is generally not zero-mean. This underscores the generality of the score-based framework described below for monitoring nonstationarity by monitoring for changes in the local mean of the score vectors, because when characterizing stochastic microstructures, we can choose from among a wide range of parametric models for the mean function $g(\bm {x}; \bm { \theta})$ that are flexible and convenient to work with, without requiring that the model exactly represents the true distribution. In our study in Section~\ref{s_SM:results}, we found that for complex materials, linear models often provide nonstationarity analyses that are nearly as effective as those for nonlinear models like neural networks, but at a much cheaper training cost.

Based on the rationale discussed in Section~\ref{s_SM:model_ms}, define and represent nonstationarity as the change in the parameters of the conditional distribution $P(X| \bm{\mathcal{N}}(X); \bm {\theta})$. This and Equation~(\ref{eqn_SM:score_mean_zero_theory}) or~(\ref{eqn_SM:score_exp_zero_emp}) imply a general method for analyzing nonstationarity through monitoring the mean behavior of the score vectors defined in Equation~(\ref{eqn_SM:score_func}). More specifically, as shown in~\cite{zhang2020concept} for monitoring for temporal nonstationarity, under fairly general conditions, when the true parameter vector changes from $\bm { \theta}$ to a different vector $\bm { \theta}'$, the expected score vector in Equation~(\ref{eqn_SM:score_mean_zero_theory}) or~(\ref{eqn_SM:score_exp_zero_emp}) differs from zero, i.e., $E_{\bm {\theta}'} [\bm {s}(\bm {\theta}; X, \bm{\mathcal{N}}(X))] \neq \bm {0}$ or $\hat{E}_{\bm { \theta}'}[\bm{s}(\hat{\bm{\theta}};X, \bm{\mathcal{N}}(X))] \neq \bm {0}$. Since our formulation equates $\bm { \theta}' \neq \bm { \theta}$ in the vicinity of a pixel with the microstructure in the vicinity of that pixel having a different distribution than that of the training/reference microstructure, we can monitor and visualize the mean of the score vector (details for which are provided in Sections~\ref{s_SM:supervised_moni} and~\ref{s_SM:unsupervised_chara}) to signal when statistically different microstructures are encountered and specifically where those differences are.

\subsection{Training the Model and Estimating the Local Mean of the Score Vector}
\label{ss_SM:spa_ave_score}
\begin{figure}[!htbp]
\centering
 \begin{subfigure}[t]{0.31\linewidth}
         \centering
        \includegraphics[width = \linewidth, trim=0in 0.03in 0in 0in, clip]{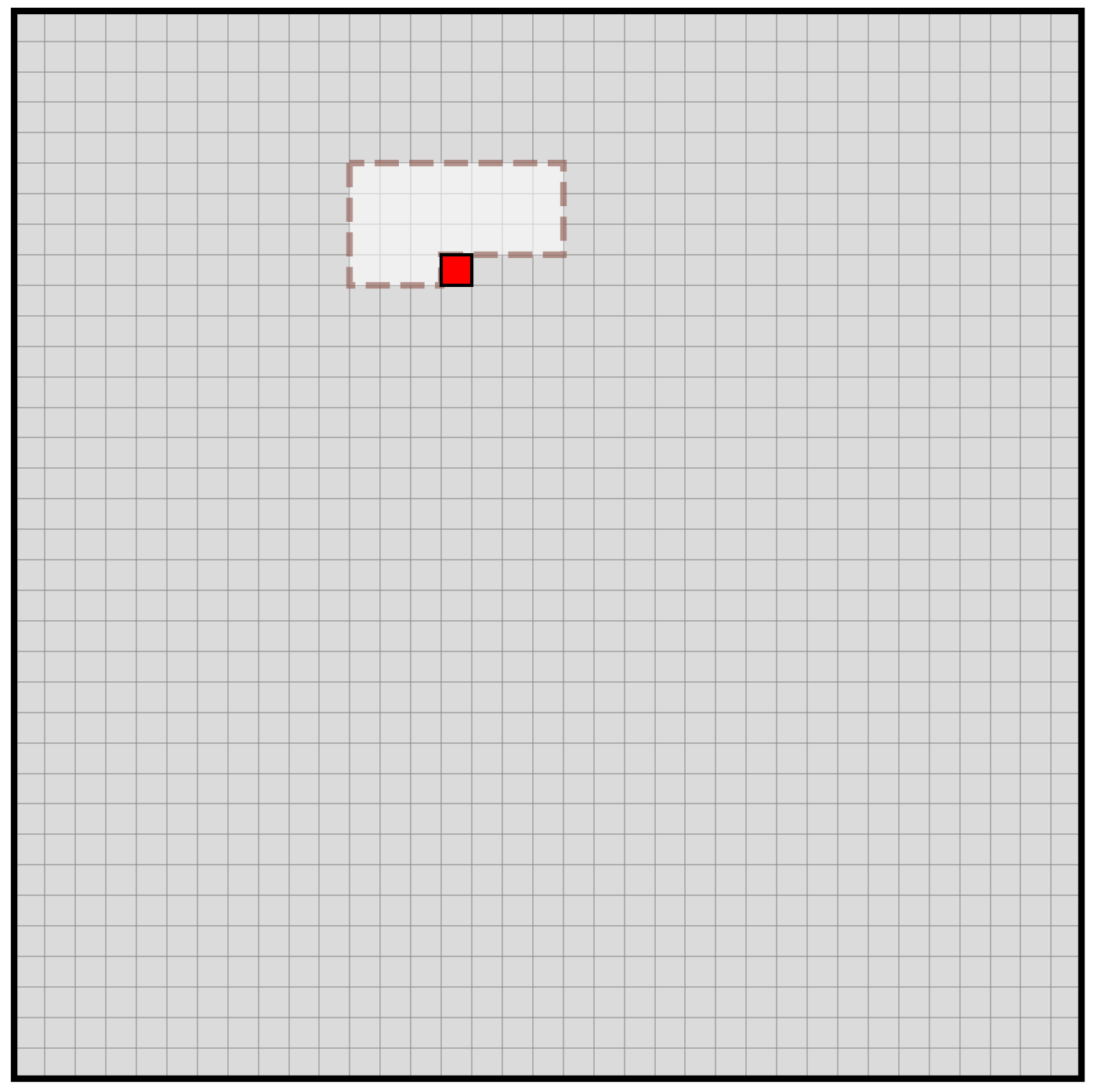}
         \captionsetup{width=.95\linewidth}
	\caption{Causal neighborhood window (excluding the red target pixel) for modeling the mean of $P(X|\bm{\mathcal{N}}(X))$, i.e., $g(\bm {x}; \bm { \theta})$.}
         \label{fig_SM:causal_wind}
  \end{subfigure}
   \begin{subfigure}[t]{0.312\linewidth}
         \centering
         \includegraphics[width = \linewidth, trim=0in 0.05in 0in 0in, clip]{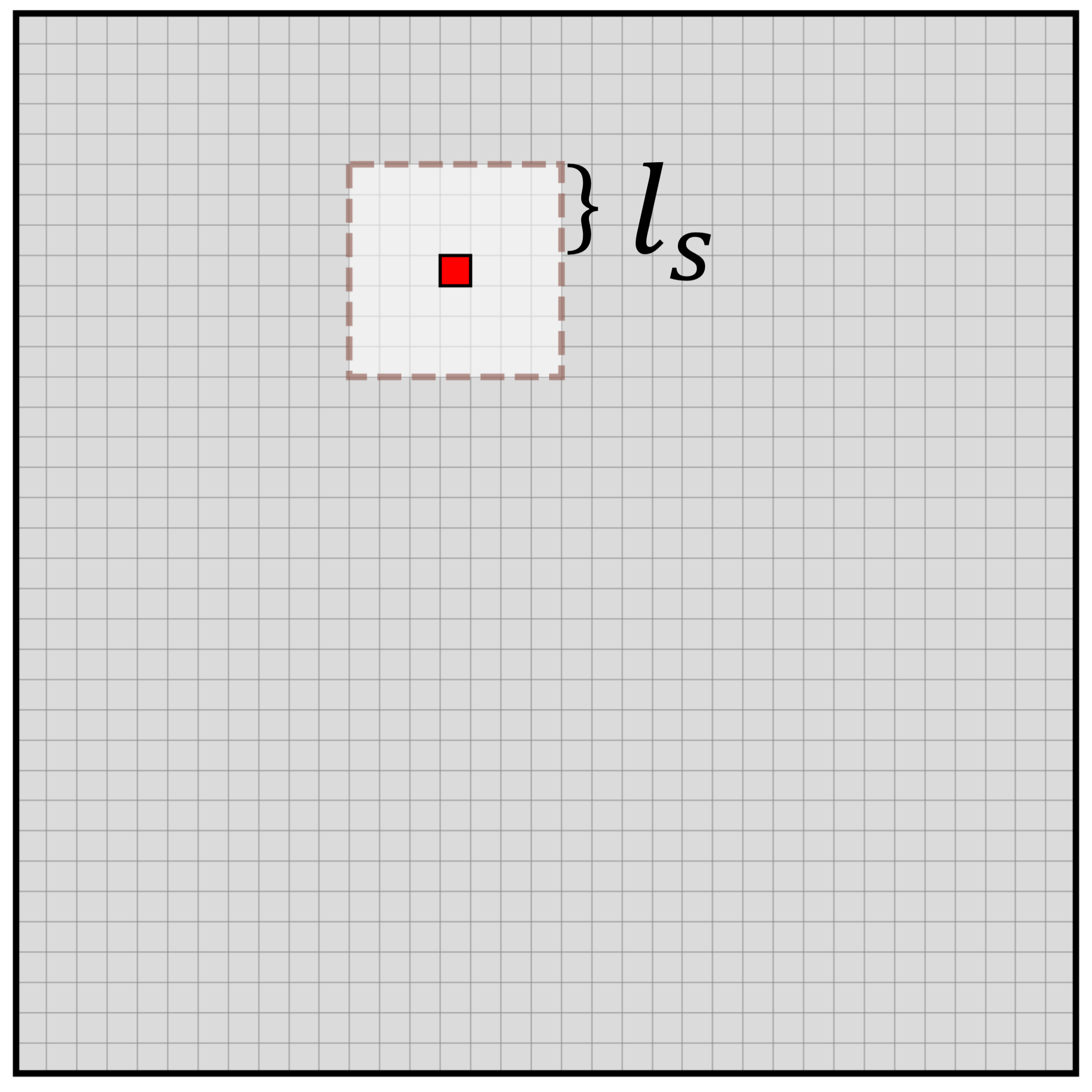}
         \captionsetup{width=.95\linewidth}
	\caption{Non-causal neighborhood window (excluding the red target pixel) with the side length as $2l_s+1$ pixels for modeling the mean of $P(X|\bm{\mathcal{N}}(X))$, i.e., $g(\bm {x}; \bm { \theta})$.}
         \label{fig_SM:non_causal_wind}
  \end{subfigure}
  \begin{subfigure}[t]{0.312\linewidth}
         \centering
         \includegraphics[width = \linewidth, trim=0in 0in 0in 0in, clip]{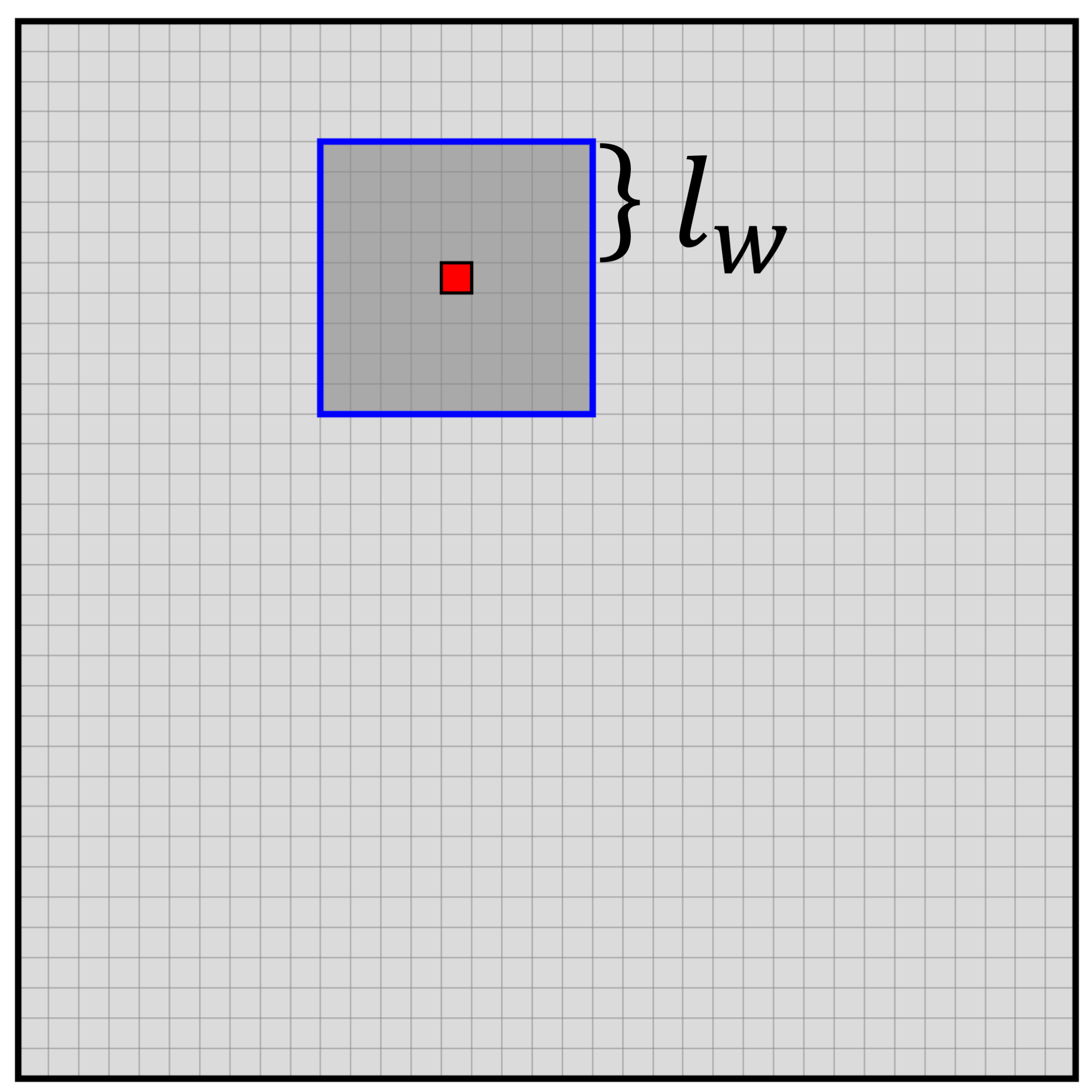}
         \captionsetup{width=.95\linewidth}
	\caption{WMA window (including the red target pixel) with the side length as $2l_w+1$ pixels for spatially smoothing score vectors.}
         \label{fig_SM:WMA_wind}
  \end{subfigure}
    \begin{subfigure}[t]{.7\linewidth}
         \centering
         \includegraphics[width = 0.75\linewidth, trim=0.in 0.in 0.in .0in, clip]{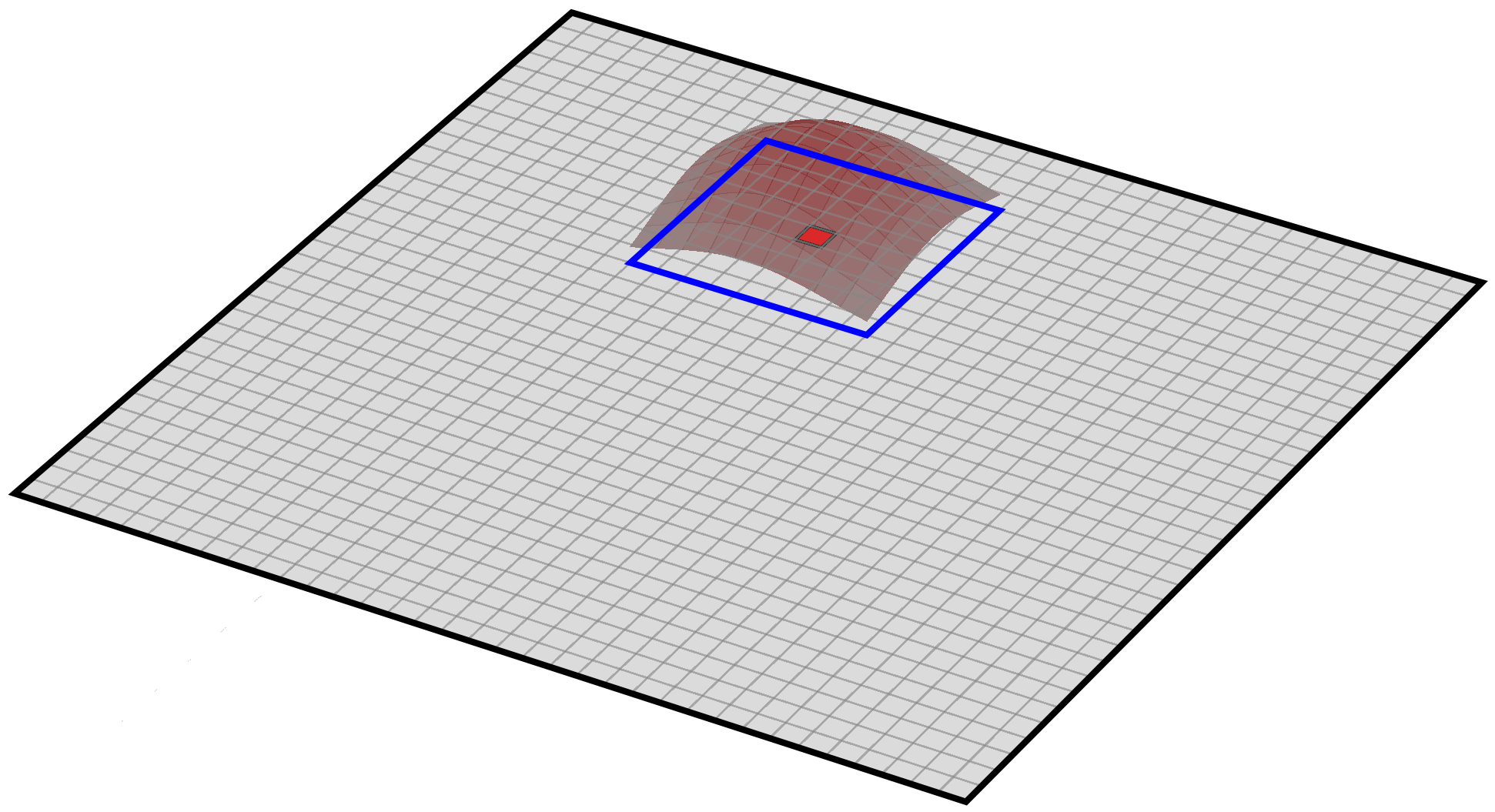}
         \captionsetup{width=.95\linewidth}
	\caption{Truncated Gaussian distribution over the WMA window.}
         \label{fig_SM:gau_dist_WMA}
  \end{subfigure}
\caption{Illustration of various windows used in computing and monitoring the score vectors. (a) Causal neighborhood window and (b) Non-causal neighborhood window (patches with brown dashed edges) for modeling the conditional distribution of $X_i$ (the red pixel) given the neighboring pixels in those windows. (c) WMA window for spatially smoothing the score vectors to estimate their local mean. To differentiate it from the neighborhood window, we use blue solid lines as the edges here. (d) A $2$D Gaussian distribution centered (at the red pixel) and truncated over the WMA window. The height is proportional to the density value of the truncated distribution.}
\label{fig_SM:causal_non_causal_wind}
\end{figure}

To train a supervised learning model $g(\bm {x}; \bm { \theta})$ to represent microstructures, we first need to choose the neighborhood $\bm{\mathcal{N}}(X)$ for each pixel $X$. As shown in Figures~\ref{fig_SM:causal_wind} and~\ref{fig_SM:non_causal_wind}, there are two basic choices: a causal or a non-causal neighborhood window (excluding the target pixel colored as red). For stationary microstructures with the Markov locality property, either a causal or a non-causal neighborhood window can serve to implicitly characterize the joint distribution $P(X)$ of the pixels in the micrograph (the former via the decomposition $P(X) = P(X_1)P(X_2|X_1)P(X3|X_1,X_2)\cdots P(X_m|X_1,\cdots,X_{m-1})$, and the latter via a mechanism akin to Gibbs sampling). In general, the choice depends on the purpose of using the trained model. For computational reasons, in certain applications the causal neighborhood window is usually chosen if the goal is to generate new samples of microstructures with the trained model~(\cite{bostanabad2016stochastic}). Also for some specific applications, a causal neighborhood window may result in better performance than a non-causal neighborhood window~(\cite{bui2018monitoringa,bui2018monitoringb}). For our purposes within our score-based monitoring framework, we choose a non-causal neighborhood window, because it obviously provides more accurate prediction of the pixel values, and it resulted in better monitoring performance with roughly the same computational cost for our usage. Intuitively, a pixel $X$ should depend on its neighbors in all directions, which suggests a non-causal neighborhood window will lead to more accurate prediction of $X$. Using the trained parametric likelihood (based on the trained supervised learning model $g(\bm{x},\bm{\theta})$), the score vectors $\bm{s}(\hat{\bm{\theta}};X, \bm{\mathcal{N}}(X))$ can be calculated for any set of image pixels. In SGD algorithms, the score vectors are by-products produced during training or predicting (e.g., for training data, $\bm {s}(\hat { \bm { \theta}}; y_i, \bm { x}_i) = \nabla _{\bm { \theta}} \log(P(X_i=y_i| \bm {\mathcal{N}}(X_i) = \bm {x}_i ; \hat{\bm { \theta}}))$ and are automatically computed as the gradient for $(y_i, \bm{x}_i)$), which is a main reason why our score-based approach is computationally reasonable.

After obtaining the score vectors, directly monitoring the individual score vectors would be an ineffective way to monitor mean changes, because the individual vectors are noisy. To handle the noise issue and estimate the local mean of the score vectors as it varies spatially across the image sample, we employ ideas similar to multivariate exponentially weighted moving average (EWMA) control chart concepts. A multivariate EWMA is of the most effective methods for monitoring for changes in the mean of general random vectors~(\cite{montgomery2007introduction}) over temporal or spatial domains. Monitoring for spatial nonstationarity (within or across samples) reduces to monitoring for changes in the mean of score vectors over regions of one sample or across multiple samples.

Similar to the temporal EWMA, we calculate the spatial weighted moving average (WMA) of the score vectors $\bm{s}(\hat{\bm{\theta}}; y, \bm {x})$ to smooth out noise and estimate the local mean, where $\hat{\bm{\theta}}$ is the MLE of the parameters computed over the entire training data. Specifically, we calculate the weighted average of the score vectors of pixels in a WMA window that is shown as the square with blue edges in Figure~\ref{fig_SM:WMA_wind}, with the weight function being a truncated $2$D normal distribution centered at the target pixel over the WMA window (including the target pixel colored as red) as illustrated in Figure~\ref{fig_SM:gau_dist_WMA}. The standard deviation $\sigma_w$ (defined below) of this truncated normal distribution governs how fast this dependency decays away from a target pixel in different directions. Notice that to distinguish the two windows, one for training a model for conditional distribution (neighborhood window), $\bm{\mathcal{N}}(X)$, and the other for WMA smoothing, we use different colors, shadings, and notations in Figure~\ref{fig_SM:causal_non_causal_wind}.

To mathematically define the WMA, let $\bm{z}_{r_i,c_i}$ denote the score vector at a pixel $X_i$ smoothed by the WMA window with row and column coordinates of $X_i$ denoted by $(r_i,c_i)$; and use $(r,c)$ as the row and column coordinates of neighboring pixels over which we calculate $\bm{z}_{r_i,c_i}$. For notational simplicity, also define $\bm{s}_{r_i, c_i} \vcentcolon= \bm{s}(\hat{\bm{\theta}}; X_i, \bm{\mathcal{N}}(X_i))$, and denote the truncated bivariate Gaussian density function by $p(r,c; (r_i,c_i, \sigma_w))$, which is viewed as a function of $(r,c)$ with the distribution centered at $(r_i,c_i)$ and having bivariate covariance matrix $\sigma_w^2\mathbf{I}$ with $\mathbf{I}$ the $2\times 2$ identity matrix. Our score vector WMA is defined as
\begin{align}
\bm {z}_{r_i,c_i} = \sum_{r=r_i-l_w}^{r_i+l_w}\sum_{c=c_i-l_w}^{c_i+l_w} p(r,c; (r_i,c_i, \sigma_w))\bm {s}_{r,c}
\label{eqn_SM:WMA_score}
\end{align}
where the window size is $2l_w+1$ as shown in Figure~\ref{fig_SM:WMA_wind}. Note that the weights $p(r,c;(r_i,c_i,\sigma_w))$ sum to one over the WMA window, by definition of the truncated distribution. In all of our examples, we use $l_w = \sigma_w$. In general, one should choose the hyper-parameter $\sigma_w$ to be approximately the size of features in the micrographs. If $\sigma_w$ is too large, fine-scale local nonstationarity will be averaged out; while if $ \sigma_w$ is too small, $\bm{z}_{r_i,c_i}$ will be too noisy.

\section{Nonstationarity Monitoring (NM)}
\label{s_SM:supervised_moni}
The goal of this monitoring is to determine whether a given set of micrograph samples are statistically equivalent to a given reference sample or samples, where, for example, the reference sample(s) represent normal conditions; and to potentially provide some interpretation if nonstationarity is detected. One practical problem falling into this category in materials fabrication or manufacturing is when we have available image samples from batches of materials that were produced under normal or well-calibrated conditions, and the goal is to determine whether subsequently produced batches of materials are statistically equivalent to the reference batch(es), based on images samples collected periodically from the subsequent batches. 

In order to monitor for local changes in the mean of the score vectors via a control chart, we use a Hotelling $T^2$~(\cite{lowry1992multivariate,hotelling1947multivariate}) statistic for $\bm{z}_{r_i,c_i}$:
\begin{align}
T^2_i = (\bm{z}_{r_i,c_i}-\bar{\bm{s}})^T\widehat{\bm{\Sigma}}^{-1}(\bm{z}_{r_i,c_i}-\bar{\bm{s}})
\label{eqn_SM:hotelling_t2}
\end{align}
where $\bar{\bm{s}}$ and $\widehat{\bm{\Sigma}}$ are the empirical mean vector and covariance matrix of the training score vectors $\{\bm{s}_{r_i,c_i}\} _{i=1} ^{m}$ over the entire training data, respectively. Notice that we first apply the WMA smoothing window and then calculate the Hotelling $T^2$ statistic for $\bm{z}_{r_i,c_i}$ as opposed to the Hotelling $T^2$ statistic for the individual score vector $\bm{s}_{r_i,c_i}$. This reduces the adverse effects of noise and gives a more accurate estimate of the local score vector mean, so that, with a proper choice of moving window size, the control limits (defined below) are tighter and the control chart is more sensitive to the mean change in score vectors. 

In the supervised learning model $g(\bm{x};\bm{\theta})$, if we view the residual standard deviation parameter $\sigma$ as an additional parameter along with $\bm{\theta}$, then changes in $\bm{\theta}$ and/or $\sigma$ indicate nonstationarity of the microstructure. Consequently, we compute and monitor the score vector with respect to both $\bm{\theta}$ and $\sigma$. Since changes in $\bm{\theta}$ versus changes in $\sigma$ represent different types of changes in the microstructure, we have found it more effective to compute a score vector for $\bm{\theta}$ and a score vector for $\sigma$ and treat them jointly but distinctly, as described below. This is relative to computing the $T^2$ statistic for the score vector for $(\bm{\theta}, \sigma)$ together. 

As mentioned above, we approximate the conditional distribution of the $i^{\mathrm{th}}$ pixel value $X_i$ given its neighboring pixels $\bm { \mathcal{N}}(X_i)$ as a normal distribution:

\begin{align}
\begin{aligned}
&P(X_i = y| \bm { \mathcal{N}}(X_i) = \bm {x}; \bm {\theta}, \sigma) = \frac{1}{\sqrt{2\pi}\sigma}\exp \left\{- \frac{[y-g(\bm {x}; \bm { \theta})]^2}{2 \sigma^2} \right\} \\
&l(\bm {\theta}, \sigma; y,\bm {x}) \vcentcolon =  \log(P(y|\bm {x}; \bm { \theta}, \sigma^2))  = - \log(\sigma) - \frac{[y-g(\bm {x}; \bm { \theta})]^2}{2 \sigma^2} + Const.
\label{eqn:log_likelihood}
\end{aligned}
\end{align}

The score vectors for $\bm {\theta}$ and $\sigma$, evaluated at the MLEs $\hat{\bm{\theta}}$ and $\hat{\sigma}$ computed from the entire training data $\{y_i, \bm{x}\}_{i=1}^m$, are defined as
\begin{align}
\bm {s}_{\bm {\theta},i} \vcentcolon =& \left.\frac{\partial l(\bm {\theta}, \sigma; y_i,\bm {x}_i)}{\partial \bm {\theta}} \right|_{(\bm{\theta}, \sigma) = (\hat{\bm{\theta}}, \hat{\sigma})} = \frac{1}{\hat{\sigma}^2} [y_i-g(\bm {x}_i; \hat{\bm { \theta}})] \nabla_{\bm {\theta}}g(\bm {x}_i; \hat{\bm { \theta}}) \label{eqn:score_theta} \\ 
 {s}_{\sigma,i} \vcentcolon =& \left.\frac{\partial l(\bm {\theta}, \sigma; y_i,\bm {x}_i)}{\partial \sigma} \right|_{(\bm{\theta}, \sigma) = (\hat{\bm{\theta}}, \hat{\sigma})}=  -\frac{1}{\hat{\sigma}}+ \frac{[y_i-g(\bm {x}_i; \hat{\bm { \theta}})]^2}{ \hat{\sigma}^3}.
\label{eqn:score_sigma}
\end{align}
Since a fixed $\hat{\sigma}$ in Equations~(\ref{eqn:score_theta}) and~(\ref{eqn:score_sigma}) drops out of the $T^2$ statistic in Equation~(\ref{eqn_SM:hotelling_t2}), it can be ignored, in which case monitoring for changes in the means of $\bm {s}_{\bm {\theta}}$ and ${s} _{\sigma}$ reduces to monitoring for changes in the mean of $[y-g(\bm {x}; \bm { \theta})] \nabla_{\bm {\theta}}g(\bm {x}; \bm { \theta})$ and changes in the mean of $[y-g(\bm {x}; \bm { \theta})]^2$. 

The goal is to detect local changes in the mean of $\bm {s} _{\bm { \theta},i}$ and/or the mean of $ s _{ \sigma,i}$. Analogous to Equations~(\ref{eqn_SM:WMA_score}) and~(\ref{eqn_SM:hotelling_t2}), we define the WMAs 
\begin{align}
\bm {z}_{\bm { \theta},r_i,c_i} =& \sum_{r=r_i-l_w}^{r_i+l_w}\sum_{c=c_i-l_w}^{c_i+l_w} p(r,c; (r_i,c_i, \sigma_w))\bm {s}_{\bm { \theta},r,c} \\
{z}_{ \sigma, r_i,c_i} =& \sum_{r=r_i-l_w}^{r_i+l_w}\sum_{c=c_i-l_w}^{c_i+l_w} p(r,c; (r_i,c_i, \sigma_w)){s}_{ \sigma,r,c}
\end{align}
and the $T^2$ statistics
\begin{align}
T^2_{\bm { \theta},i} = (\bm{z}_{\bm { \theta},r_i,c_i}-\bar{\bm{s}} _{\bm { \theta}})^T\widehat{\bm{\Sigma}}_{\bm { \theta}}^{-1}(\bm{z}_{\bm {\theta},r_i,c_i}-\bar{\bm{s}}_{\bm{\theta}}),
\end{align}
where $\bar{\bm{s}} _{\bm { \theta}}$ and $\widehat{\bm{\Sigma}}_{\bm { \theta}}$ are the empirical mean vector and covariance matrix of the training score vectors $\{\bm{s}_{\bm { \theta},r_i,c_i}\} _{i=1} ^{m}$. Because $\bm {s} _{\bm { \theta},i}$ is a vector and $T^2 _{\bm { \theta},i}$ is intended to detect a mean change in $\bm {s} _{\bm { \theta},i}$ away from $\bar {\bm {s}} _{\bm { \theta}}$ in any direction in the $\bm { \theta}$ space, the control chart for $T^2_{\bm{\theta},i}$ has only an upper control limit ($UCL_{\bm{\theta}}$), the purpose of which is to signal a change if $T^2_{\bm{\theta},i} > UCL_{\bm{\theta}}$.  Since $s_{\sigma,i}$ is a scalar, its WMA chart detects a change in its mean if either $z_{\sigma,r_i,c_i} < LCL_{\sigma}$ or $z_{\sigma,r_i,c_i} > UCL_{\sigma}$, where $LCL_{\sigma}$ and $UCL_{\sigma}$ denote its lower and upper control limits, respectively. 

We use a multi-chart to monitor for mean changes in either $\bm{s}_{\bm{\theta},i}$ or $s_{\sigma,i}$. A change in either in the vicinity of pixel $X_i$ indicates that the stochastic nature of the microstructure has changed in the vicinity of this pixel, relative to the reference microstructure samples. The multi-chart is defined as signaling either if $T^2_{\bm{\theta},i} > UCL_{\bm{\theta}}$ or if $z_{\sigma,r_i,c_i} < LCL_{\sigma}$ or $z_{\sigma,r_i,c_i} > UCL_{\sigma}$. To display the control chart results in a single $3$D plot (e.g., as in Figure~\ref{fig_SM:2d_ar_sup_moni_multi_chart_nnet}, later), we define the plotted scaled statistics from the two component charts as

\begin{align}
{C}_{{\bm { \theta}}, i} =& \frac{2T^2_{{\bm { \theta}},i}}{UCL_{{\bm { \theta}}}}-1\\
{C} _{{\sigma}, i} =& \frac{z_{\sigma,r_i,c_i} - \frac{UCL_{{\sigma}}+LCL_{{\sigma}}}{2}}{\frac{UCL_{{\sigma}}-LCL_{{\sigma}}}{2}},
\end{align}
which would individually signal if either $C_{\bm{\theta},i} > 1$ or $|C_{\bm{\theta},i}| > 1$. The plotted statistic in the multi-chart statistic is then defined as
\begin{align}
C_{M,i} = \text{sign}({C}_{{\bm { \theta}},i} + {C} _{{\sigma},i}) \times \max(|{C}_{{\bm { \theta}},i}|, |{C} _{{\sigma},i}|),
\end{align}
and the multi-chart signals nonstationarity in the vicinity of pixel $X_i$ if $C_{M,i}$ falls outside the range $[-1,1]$. 

The three control limits ($UCL_{\bm{\theta}}$, $LCL_{\sigma}$, and $UCL_{\sigma}$) are determined empirically, as follows. We divide the reference data $\{(y_i, \bm{x}_i)\} _{i=1} ^{M}$ into two sets $\{(y_i, \bm{x}_i)\} _{i=1} ^{m}$ (which we refer to as the training data) and $\{(y_i, \bm{x}_i)\} _{i=m+1} ^{M}$ (which we refer to as the CL-selection data).  We fit the predictive model $g(\bm{x};\bm{\theta})$ to the training data, and then for each pixel $X_i$ in the CL-selection data we compute the chart statistics $\{T^2_{\bm{\theta},i}, z_{\sigma,r_i,c_i}\} _{i=m+1} ^{M}$.  We then compute the empirical cdfs of $\{T^2_{\bm{\theta},i}\}_{i=m+1}^{M}$ and $\{z_{{ \sigma},r_i,c_i}\}_{i=m+1}^{M}$.
The $UCL_{{\bm { \theta}}}$, $LCL_{{\sigma}}$, and $UCL_{{\sigma}}$ are determined such that (1) the two component charts for $C_{\bm{\theta},i}$ and $C_{{\sigma},i}$ each have the same empirical false alarm rate over the CL-selection data, and (2) the multi-chart has an empirical false alarm rate over the CL-selection data that is equal to some user-specified desired false alarm rate. The three control limit values can be efficiently found via searching for the required empirical false alarm rate (denoted by $\alpha_{\bm { \theta}, \sigma}$) over the CL-selection data for the two component charts (i.e., for $\bm {s} _{\bm { \theta}}$ and $s _{ \sigma}$ individually) so that the empirical false alarm rate (denoted by $\alpha_M$) of the multi-chart over the CL-selection data is the desired value. This can be accomplished via binary search algorithm by noting that $ \alpha _{M}$ is monotonically non-decreasing with $ \alpha _{\bm { \theta}, \sigma}$. In the sequel, we denote these two component charts and the multi-chart by as SWMA-$\bm { \theta}$, SWMA-$ \sigma$, and SWMA-M, where the ``S" stands for ``score-based".

Selecting a benchmark method to which to compare our score-based approach is difficult, because there are very few existing methods that have been developed to monitor for nonstationarity on a pixel-by-pixel basis. To the best of our knowledge, the closest existing method is the residual-based method in~\cite{bui2018monitoringb}. Although it was more intended to detect local defects in the images, it can easily be adapted to monitor for nonstationarity on a pixel-by-pixel basis. To adapt the method for this purpose, we monitor the residuals (i.e., prediction errors) $r_i = X_i - g(\bm{\mathcal{N}}(X_i);\hat{\bm{\theta}})$ for each pixel. The control charts for residuals (which we refer to as the RWMA) have both a $LCL$ and an $UCL$, which we choose based on the empirical distribution of the residuals over the same set of CL-selection data, similarly to how we determine the control limits for the score-based charts.

The following summarizes the steps of the NM approach.

\begin{itemize}
\item
\textbf{Step $1$~(Training)}: The data $\{(y_i, \bm{x}_i)\}_{i=1}^M$ obtained from reference micrograph/micrographs is split into the two subsets $ \{(y_i, \bm{x}_i)\} _{i=1} ^{m}$ and \\$\{(y_i, \bm{x}_i)\} _{i=m+1} ^{M}$. The first subset $\{(y_i, \bm{x}_i)\} _{i=1} ^{m}$ is used to train the parametric supervised learning model $g(\bm {x}; \bm { \theta})$, mean of the conditional likelihood $P(y|\bm{x};\hat{\bm{\theta}})$ of the individual pixels $(y,\bm{x})$, which implicitly represents the underlying joint distribution of the micrograph pixels. During training, we compute the (regularized) MLE $\hat {\bm { \theta}}$ by minimizing the cost function, $-\sum_{i=1} ^{m} l(\bm { \theta}, \sigma; y_i, \bm {x}_i) + J(\bm { \theta})$, where $J(\bm { \theta})$ is a regularization term (in this study we use $J(\bm { \theta})= \lambda||\bm { \theta}||_{L_2}^2$) with all  hyper-parameters chosen by cross-validation. As discussed in~\cite{zhang2020concept}, including regularization in the training loss function does not affect the salient point that the mean of the score functions changes if and only if the predictive relationship is nonstationary.
\item
\textbf{Step $2$~(CL-selection)}: The supervised learning model from Step $1$ is applied to the second subset $\{(y_i, \bm{x}_i)\} _{i=m+1} ^{M}$ to obtain the score vectors $\{s_{\bm{\theta},i}, s_{\sigma,i}\}_{i=m+1}^M$ (and the residuals $\{r_i\}_{i=m+1}^M$ for the residual-based benchmark) and to select the control limits as described above to provide an empirical false alarm rate that is equal to some user-specified desired false alarm rate $\alpha$ (e.g., $0.01$ or $0.001$).
\item
\textbf{Step $3$~(Monitoring)}: New micrograph samples are collected and, converted to a monitoring data set $\{(y_i, \bm {x}_i)\} _{i > M}$, for each pixel of which the same score vectors, residuals, and monitoring statistics $\{C_{\bm{\theta},i}, C_{\sigma,i}, C_{M,i}, r_i\}$ are computed and compared to their respective control limits. For a particular chart, if a significant portion of the charted statistics fall beyond its control limit, we conclude that the new sample is nonstationary in the sense that its microstructure differs stochastically from the reference sample.
\end{itemize}
Before fitting the predictive model $g(\bm{x};\bm{\theta})$ to the training data and applying it to the CL-selection or monitoring image data to compute the score vectors or residuals, we always scale each micrograph to have zero mean and unit variance, because a change in the mean or variance of the pixel grayscale values across different micrographs could be due to the different light exposure or contrast levels, which should not be treated as nonstationarity. If one suspects that such mean or variance changes could be the net result of actual nonstationarity of the microstructure, then one could supplement our score-based multi-chart with additional component charts that monitor a local WMA estimate of the pixel mean and variance. For the monitoring micrograph(s) (i.e., the new set of micrographs to be monitored and compared to the reference micrographs), we define the \textit{power} (i.e., probability of correctly detecting nonstationarity) for a control chart as the percentage of pixels in the monitoring samples for which the control chart signaled, i.e., for which the control chart statistic fell outside its control limit(s). The larger the power, the more clearly the control chart correctly indicated nonstationarity, when the monitoring micrographs are truly nonstationary. In Section~\ref{ss_SM:res_supervised_moni}, we apply and demonstrate this NM approach on a real and simulated materials data.

\section{Nonstationarity Diagnostics (ND)}
\label{s_SM:unsupervised_chara}
The NM approach in Section~\ref{s_SM:supervised_moni} is intended to (1) indicate whether the samples are nonstationary in the sense that a monitoring sample or part of a monitoring sample is different than the CL-selection reference sample(s); and (2) if nonstationary, highlight regions that are most likely to be different. If there are multiple phases in a nonstationary sample (or samples), a related important objective is to identify and demark the regions of the sample that correspond to the different phases (i.e., different distinct types of microstructure stochastic behavior). We refer to this objective as nonstationarity diagnostics (ND) to distinguish it from the NM objective. In this section, we develop an approach for this that can be used as a follow-up to NM, e.g., if the NM indicates a sample is nonstationary, to determine how many material phases there are and, more generally, what is the nature of the nonstationarity. Alternatively, our ND approach can be used as a stand-alone approach in which we are given a single sample or multiple samples and want to know the nature of the nonstationarity. For example, suppose a materials scientist has just created a new sample of material in the laboratory using some new processing technique or settings, and one goal is to understand the nature of the new material and whether there are multiple material phases mixed together. Or suppose that engineers in a commercial-scale process have just implemented a new processing method and want to know the same.

The intuition behind our score-based method for ND is as follows. If a micrograph is a realization of a stationary random process, then the same value of $\bm{\theta}$ will represent the microstructure behavior everywhere in the micrograph. In this case, in addition to Equation~(\ref{eqn_SM:score_exp_zero_emp}) holding, the sample score vectors $\{\bm{s}(\hat{\bm{\theta}};y_i, \bm{x}_i)\}_{i=1}^m$ will have local empirical mean close to zero over every local region, providing the region is large enough that the noise in the score vectors averages out. In contrast, if the micrograph has multiple phases in it as in Figures~\ref{fig_SM:PMMA} and~\ref{fig_SM:dual_phase_steel}, there will be multiple $\bm{\theta}$'s, say $\{\bm{\theta}^{(l)}\}_{l=1}^k$, representing the stochastic nature of the $k$ different phases. If we hypothetically had \textit{a priori} pixel-wise phase labels to serve as ground truth for all $k$ phases in the micrograph/micrographs, we would train $k$ different supervised learning models to represent the $k$ different conditional distributions $\{P(y|\bm{x};\bm{\theta}^{(l)})\}_{l=1}^k$ for the $k$ different phases. For each phase, as in Equation~(\ref{eqn_SM:score_mean_zero_theory}) or~(\ref{eqn_SM:score_exp_zero_emp}), we would have $E_{\bm{\theta}^{(l)}}[\bm{s}(\bm{\theta}^{(l)}; X,\bm{\mathcal{N}}(X))]=\bm{0}$ or $\hat{E}_{\bm { \theta} ^{(l)}}[\bm{s}(\hat{\bm{\theta}}^{(l)};X, \bm{\mathcal{N}}(X))] = \bm {0}$, where $\hat { \bm { \theta}} ^{(l)}$ is the MLE of ${ \bm { \theta}}^{(l)}$ over its corresponding region. In reality, we do not have such labels \textit{a priori}, and even though we fit a single model to the entire training data with multiple phases present, there is no single parameter vector $\bm{\theta}$ that can represent the multiphase microstructure. However, we can still leverage the preceding concepts that distinguish from the case of a stationary microstructure to diagnose the nonstationarity in the micrographs. More specifically, we can train a single model for the entire set of training data, in which case the empirical (sample) mean of the score vectors over the entire training data is zero, i.e., $\sum_{i=1}^m \bm{s}(\hat{\bm{\theta}}; y_i,\bm{x}_i) \cong \bm{0}$, where $\hat{\bm{\theta}}$ is the MLE over the entire training data. But locally, within regions that fall inside any of the $k$ single phases, the score vectors will have nonzero mean. This is because each individual phase has different stochastic behavior than the mixture of all $k$ phases, and the latter is what $\hat{\bm{\theta}}$ represents. 

As an alternative interpretation, score vectors are the updating vectors in a SGD algorithm to fit the model to maximize the log-likelihood function, i.e., $\hat{\bm{\theta}}_{t+1} = \hat{\bm{\theta}}_{t} + \eta \bm{s}(\hat{\bm{\theta}}_{t}; y_i, \bm{x}_i)$,  at iteration $t$ of the algorithm. The score vectors from within any particular stationary phase from among the $k$ phases, say the $l^{\mathrm{th}}$ phase, will tend to steer the current parameter values $\hat{\bm{\theta}}$ towards $\bm{\theta}^{(l)}$. If the $k$ phases are sufficiently different, then $\{\bm{\theta}^{(l)}\}_{l=1}^k$ will be sufficiently different in the high-dimensional space, and so the mean vectors of the $k$ score vectors, $\bm{s}(\hat{\bm{\theta}};y^{(l)}, \bm{x}^{(l)}), l\in\{1,2,\cdots,k\}$ should generally be different and non-zero, where $(y^{(l)}, \bm{x}^{(l)})$ represents the values of a pixel and its neighborhood from within the $l^{\mathrm{th}}$ phase. Based on this intuition, our score-based ND approach uses \textit{k-means} clustering on the score vectors $\{\bm{s}(\hat{\bm{\theta}};y_i, \bm{x}_i)\}_{i=1}^m$ (or some transformed version thereof) to diagnose the spatial nonstationarity of microstructures. 

To reduce the effects of noise in the score vectors and improve the clustering performance by taking into account spatial proximity information, we conduct clustering on $\{\bm{z}_{r_i, c_i}\}_{i=1}^m$ in Equation~(\ref{eqn_SM:WMA_score}) instead of original score vectors, as follows. First, as a visualization tool to help approximately estimate the number $k$ of phases in the sample, we developed a $3$D plot in which the two horizontal axes represent the $2$D spatial coordinates of the micrograph, and the vertical axis represents the magnitude of the vectors $\{\bm{z}_{r_i, c_i}\}_{i=1}^m$. Moreover, the red-green-blue color of this plotted surface represents the first $3$ principal components analysis (PCA) scores. See Figure~\ref{fig_SM:silica_PMMA_visu}, later, as an example of this plot. In this plot, the pixels falling into different phases can be distinguished based on having different vertical axes height and/or different color, because the smoothed score vectors from different phases have different magnitudes and directions. From this plot, the estimated number $\hat {k}$ of phases is taken, roughly, to be the number of patches with different heights and different colors. This $\hat {k}$ is then used in setting the number of centroids in the \textit{k-means} clustering algorithm. A $3$D scatter plot of the top-$3$ PCA scores of $\{\bm{z}_{r_i, c_i}\}_{i=1}^m$ along with their clustering labels can be constructed, and those labels can be also overlaid on the initial micrograph to show regions of different phases. In Section~\ref{ss_SM:res_unsupervised_chara}, we demonstrate this ND approach on real and simulated materials data.

\section{Experimental Results}
\label{s_SM:results}

In this section, we present the results of our score-based framework using real and simulated micrograph data for NM (Section~\ref{ss_SM:res_supervised_moni}) and ND (Section~\ref{ss_SM:res_unsupervised_chara}).

\subsection{Results for the NM Approach}
\label{ss_SM:res_supervised_moni}
\begin{figure}[!htbp]
\centering
 \begin{subfigure}[t]{0.32\linewidth}
         \centering
         \includegraphics[width = \linewidth, trim=0in 0in 0in 0in, clip]{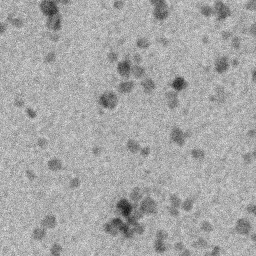}
         \captionsetup{width=.95\linewidth}
	\caption{}
         \label{fig_SM:pmma_ref}
  \end{subfigure}
   \begin{subfigure}[t]{0.32\linewidth}
         \centering
         \includegraphics[width = \linewidth, trim=0in 0.0in 0in 0in, clip]{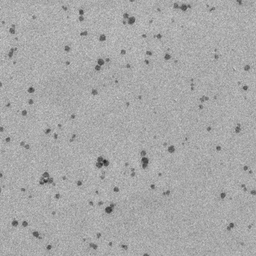}
         \captionsetup{width=.95\linewidth}
	\caption{}
         \label{fig_SM:pmma_moni}
  \end{subfigure}
\caption{Example of (a) reference and (b) subsequent nonstationary micrographs of silica particles dispersed in PMMA.}
\label{fig_SM:data_supervised_moni}
\end{figure}

We consider two data sets: a PMMA data set and a simulated $2$D AR data set shown in Figures~\ref{fig_SM:data_supervised_moni} and~\ref{fig_SM:ar_model1}-\ref{fig_SM:ar_model2}, respectively. The former is a real data set that consists of TEM images of silica particles in PMMA with octyl functional modification. How densely those particles are dispersed in the matrix can be controlled by various processing conditions, and the dispersion density can affect the physical properties~(e.g., the breakdown stress or dielectric constant of the material). We select two sets of samples with different dispersion densities and treat one as the reference data and the other as the monitoring data to be monitored for nonstationarity. The simulation data are generated via a $2$D spatial autoregressive (AR) model with various choice of AR coefficients. The $2$D AR model allows us to generate many Monte Carlo replicates of data sets to investigate and compare the false alarm rate and the power of the various control charts. We show that our score-based charts are far more effective than the residual-based chart, and its performance is further enhanced by the multi-chart.

\subsubsection{PMMA Data Analysis}
\label{sss_SM:res_supervised_moni_pmma}
\begin{figure}[!htbp]
\centering
\begin{subfigure}[t]{0.455\linewidth}
          \centering
          \includegraphics[width = \linewidth, trim=0in -1.07in 0in 0in, clip]{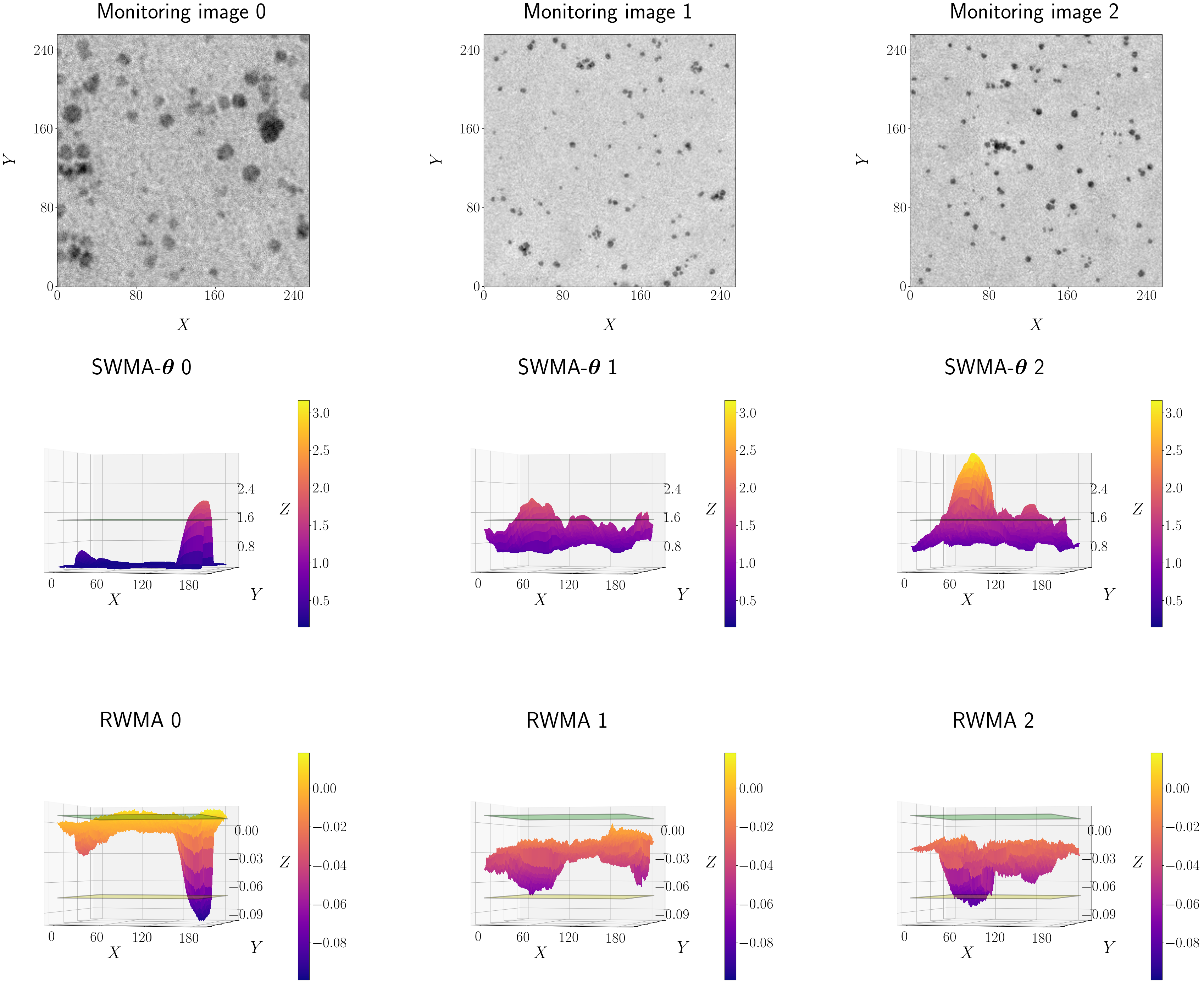}
          \captionsetup{width=.95\linewidth}
 	\caption{}
          \label{fig_SM:pmma_sup_nnet}
   \end{subfigure}\hspace{0.03\textwidth}
   \begin{subfigure}[t]{0.47\linewidth}
          \centering
          \includegraphics[width = \linewidth, trim=0in 0.2in 0in 0in, clip]{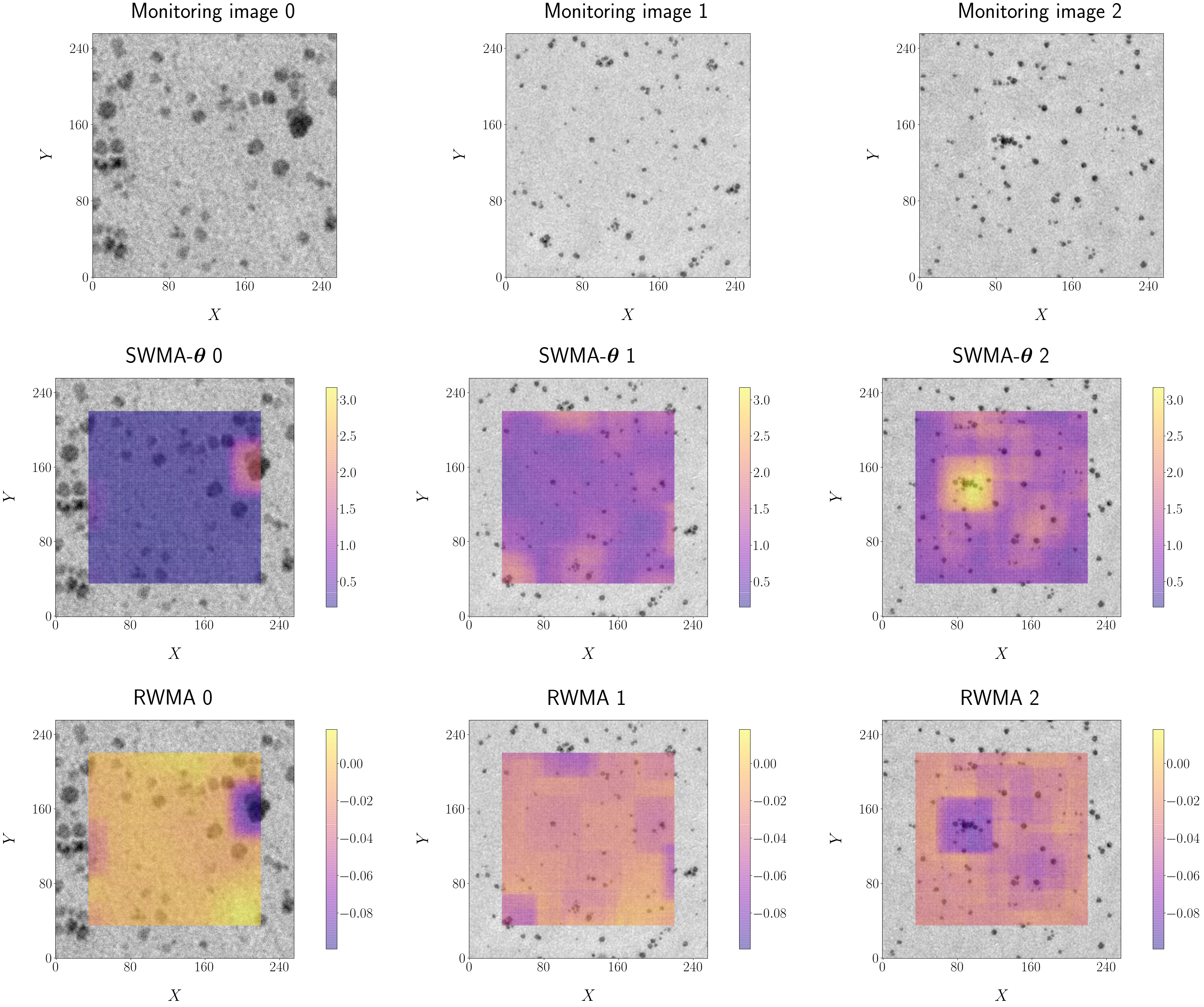}
          \captionsetup{width=.95\linewidth}
 	\caption{}
          \label{fig_SM:pmma_sup_hm_nnet}
   \end{subfigure}   
\caption{Control charts of SWMA-$\bm { \theta}$ and RWMA on the PMMA data set from the neural network model. The window length scale for training is $l_s=5$, and the WMA window length scale $l_w=30$.}
\label{fig_SM:pmma_sup_moni_nnet}
\end{figure}

We trained two different supervised learning models $g(\bm{x}; \bm{\theta})$, a linear regression model and a fully-connected neural network model with one hidden-layer having $10$ nodes, to serve as the mean function of the conditional distribution. The $L_2$ regularization parameter, $\lambda = 0.01$, was chosen via cross-validation. The results for the PMMA micrographs in Figures~\ref{fig_SM:data_supervised_moni} are shown in Figure~\ref{fig_SM:pmma_sup_moni_nnet} for the neural network model. Although the linear model results in much less accurate prediction of each pixel value, it provided similar monitoring results as the neural network and so is omitted here.

Figures~\ref{fig_SM:pmma_sup_nnet} and~\ref{fig_SM:pmma_sup_hm_nnet} show $3$D control charts and $2$D heat-maps, respectively, of the results. In the $3$D control charts shown in Figure~\ref{fig_SM:pmma_sup_nnet}, we plot the $UCL$ for the SWMA-$\bm{\theta}$ chart and the $LCL$ and $UCL$ for the RWMA chart as horizontal planes, calculated from CL-selection micrographs. For comparison, the first monitoring micrograph is a reference micrograph, and the other two monitoring micrographs are from nonstationary micrographs which are statistically different from reference ones (so we would like the chart to signal for these micrographs).

From the control charts in Figure~\ref{fig_SM:pmma_sup_nnet}, we can see that the SWMA-$\bm{\theta}$ chart out-performs the RWMA chart in terms of the power (i.e., the out-of-control percentage of signals for the two nonstationary images). For the heat-maps in Figure~\ref{fig_SM:pmma_sup_hm_nnet}, we used the following convention: (1) for all heat-maps of each monitoring statistic (each row), we apply the same color scheme, meaning the color bars correspond to the same range of numbers, the minimum and maximum values of which correspond to the two extremes of the monitoring statistic values of all the heat-maps in that row; and (2) the color bars of different rows share the same colors, but correspond to different ranges of numbers. In this way, within each row, we can map the relative differences in metric values to the differences of colors in heat-maps. And then we can compare the relative differences across rows. We observe the SWMA-$\bm { \theta}$ chart has the best performance in terms of best distinguishing the two monitoring images from the reference image. Besides the global nonstationarity, the score-based method seems to be also effective in detecting local nonstationarity, based on the peaks in the heat-maps coinciding with what appear to be somewhat unusual local agglomerations. Here, we only consider the SWMA-$\bm {\theta}$ chart but not the SWMA-$\sigma$ chart, because the latter generally requires the CL-selection data to be of larger size, as in the example in Section~\ref{sss_SM:res_supervised_moni_2d_ar}. This is perhaps because the scores for the SWMA-$\sigma$ chart are the squares of the residuals, which typically have higher variability than the scores for the SWMA-$\bm{\theta}$ chart.  

\subsubsection{$2$D AR Data Analysis}
\label{sss_SM:res_supervised_moni_2d_ar}
\begin{figure}[!htbp]
\centering
   \begin{subfigure}[t]{0.46\linewidth}
          \centering
         \includegraphics[width = \linewidth, trim=0in 0in 0in 0in, clip]{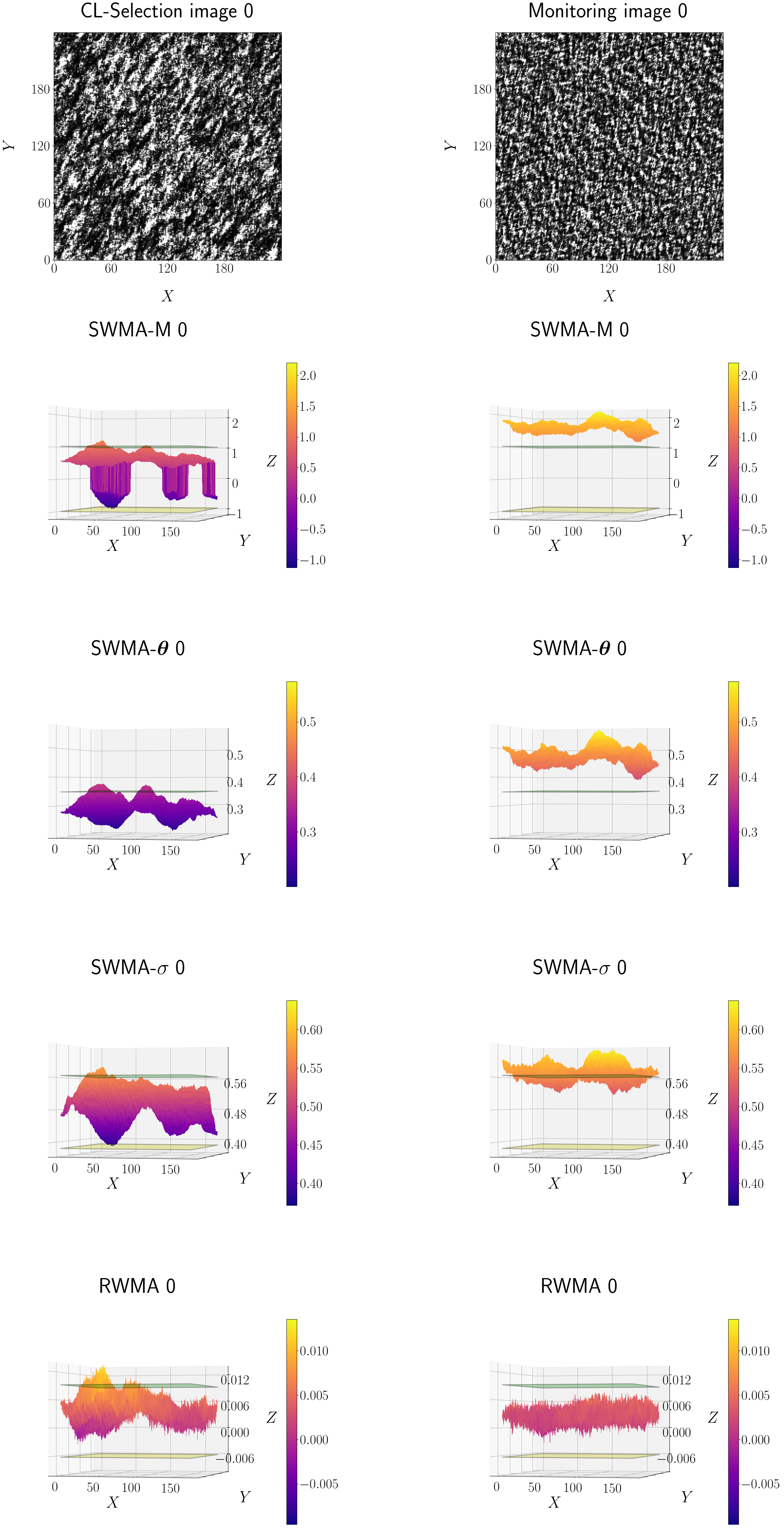}
          \captionsetup{width=.95\linewidth}
 	\caption{}
          \label{fig_SM:2d_ar_sup_multi_chart_nnet}
   \end{subfigure} \hspace{0.04\textwidth} 
   \begin{subfigure}[t]{0.46\linewidth}
          \centering
            \includegraphics[width = \linewidth, trim=0in 0in 0in 0in, clip]{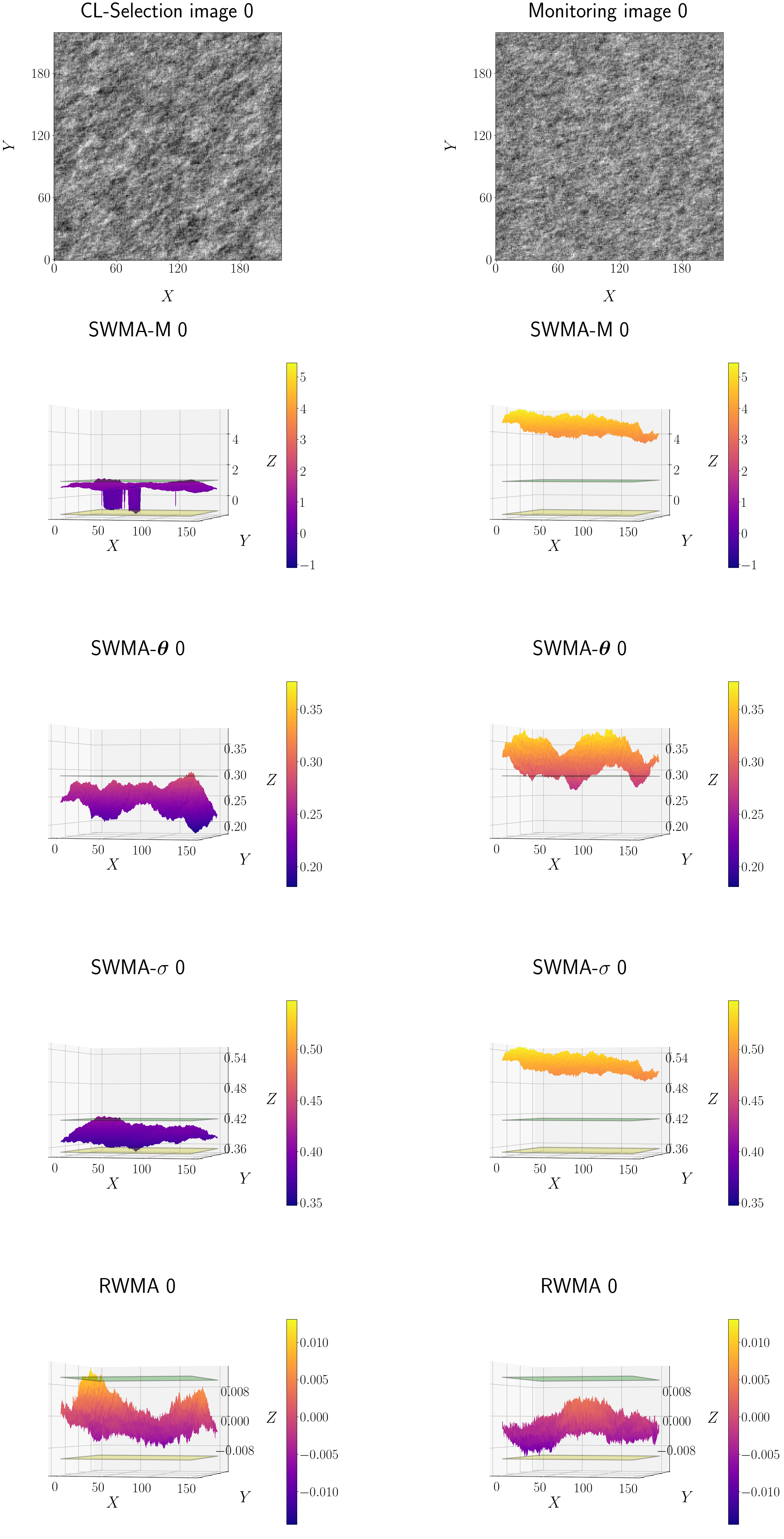}
          \captionsetup{width=.95\linewidth}
 	\caption{}
          \label{fig_SM:2d_ar_sup_multi_chart_nnet_1}
   \end{subfigure}  
\caption{Spatial WMA control chart results for the $2$D AR data. Top row shows two pairs of microstructures. The left micrograph in each pair (panel (a) or panel (b)) is representative of the reference images, and the right micrograph is statistically different and representative of the monitoring images. The training and WMA window length scales are $l_s = 5$ and $l_w = 30$.}
\label{fig_SM:2d_ar_sup_moni_multi_chart_nnet}
\end{figure}

\begin{figure}[!htbp]
\centering
 \begin{subfigure}[t]{.485\linewidth}
         \centering
         \includegraphics[width = \linewidth, trim=0in 0in 0in 0in, clip]{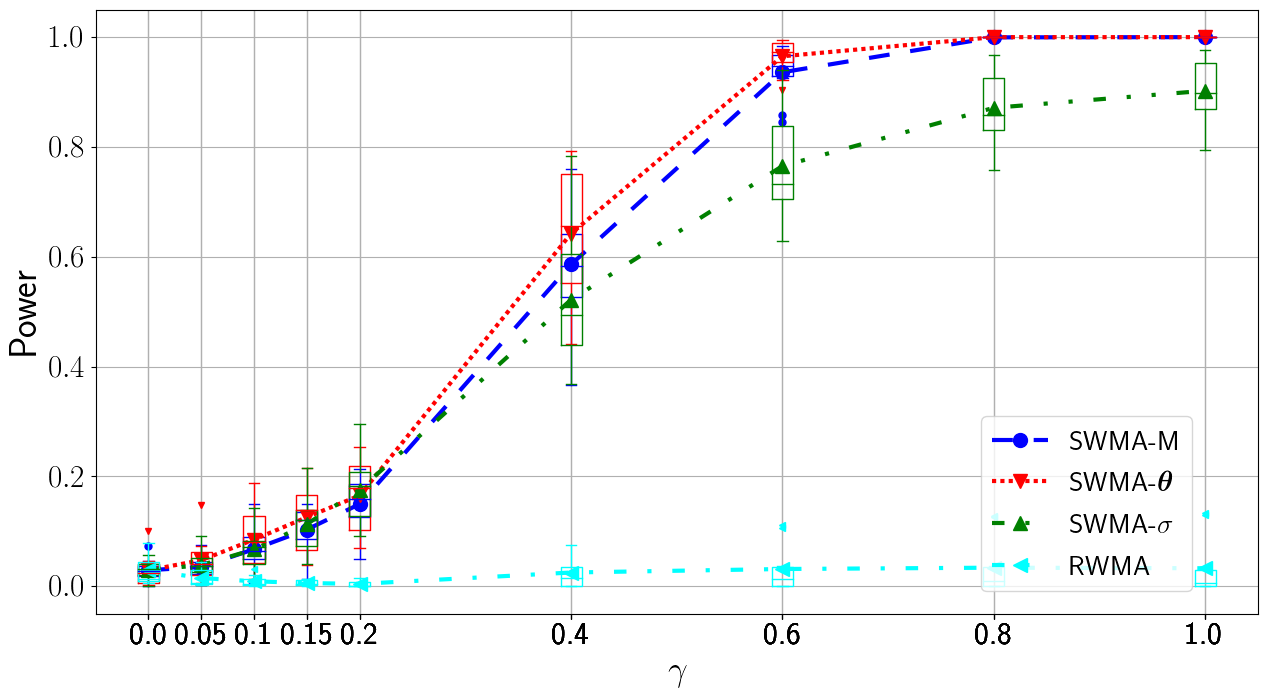}
         \captionsetup{width=.95\linewidth}
	\caption{}
         \label{fig_SM:power_comp_nnet}
  \end{subfigure} \hspace{0.01\textwidth}
  \begin{subfigure}[t]{.485\linewidth}
         \centering
         \includegraphics[width = \linewidth, trim=0in 0in 0in 0in, clip]{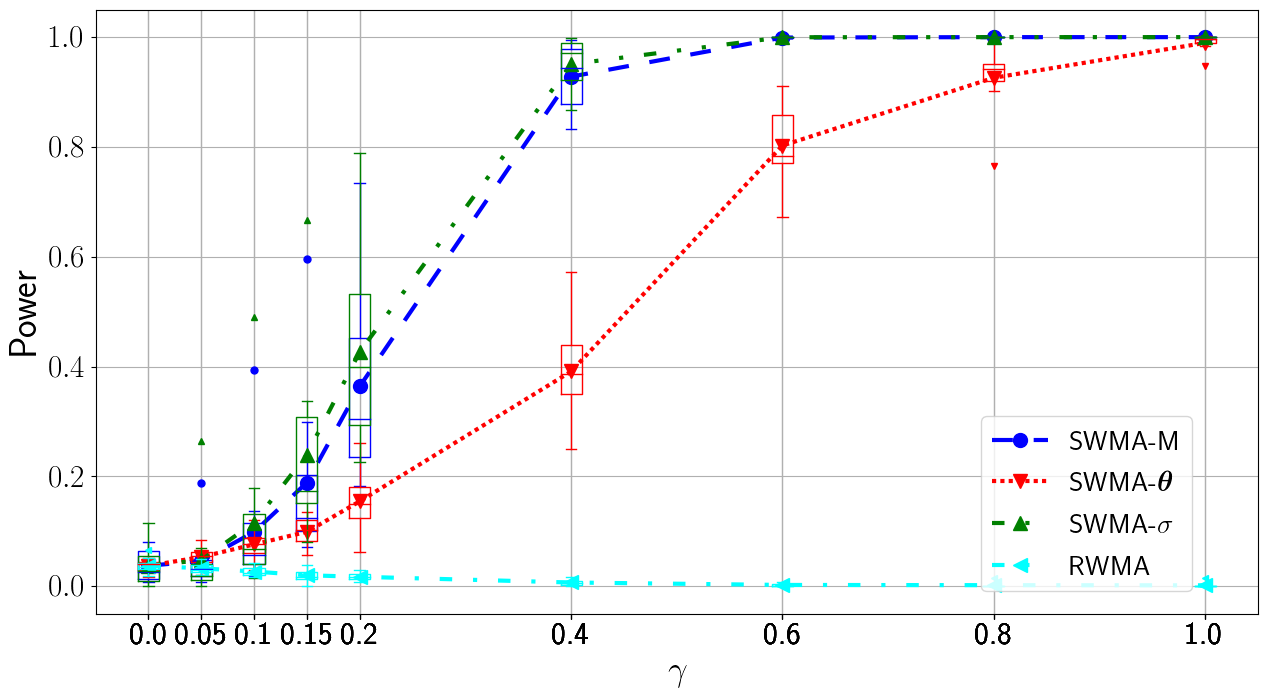}
         \captionsetup{width=.95\linewidth}
	\caption{}
         \label{fig_SM:power_comp_nnet_1}
  \end{subfigure}
\caption{Comparison of the statistical power of the four control charts in detecting nonstationarity of micrographs corresponding to the Figure~\ref{fig_SM:2d_ar_sup_moni_multi_chart_nnet} example, as $\gamma$ varies from $0$ to $1$. The power for $\gamma = 0$ is the false alarm rate. Panels (a) and (b) are for the AR models corresponding to Figures~\ref{fig_SM:2d_ar_sup_multi_chart_nnet} and~\ref{fig_SM:2d_ar_sup_multi_chart_nnet}, respectively. For each value of $\gamma$, the box plots are the power values across ten Monte Carlo replicates.}
\label{fig_SM:2d_ar_power_comp}
\end{figure}

In order to assess the power of the score-based methods on different kinds of nonstationarity, we use a $2$D AR model to generate stochastic microstructures. The model is (for $i \in \{1, 2, \cdots, m\}$):
\begin{align}
\begin{aligned}
U_{r_i, c_i} =& c_0+ \sum_{r = 0}^{l_g} \sum_{c=0}^{l_g} \phi_{r,c} U_{r_i-r, c_i-c}+ \epsilon_{r_i, c_i}\\
\epsilon_{r_i, c_i} \sim& NID(0, \sigma_{\mathrm{AR}}^2) \\
X_i = & h(U_{r_i, c_i})
\end{aligned}
\label{eqn_SM:2d_ar_data_gen}
\end{align}
where $(r_i, c_i)$ are the row and column coordinates of the $i^{\mathrm{th}}$ pixel; $c_0$ is an intercept parameter; $U_{r_i, c_i}$'s are latent variables; $l_g$ is the length scale of data generation window; $\{\phi_{r, c}\}_{r,c=0}^{l_g}$ with $\phi_{0,0}=0$ are the AR coefficients; $\sigma_{\mathrm{AR}}^2$ is the variance of i.i.d Gaussian noise random variable $\epsilon_{r_i, c_i}$; and the function $h(\cdot)$ takes a latent variable $U_{r_i, c_i}$ and outputs a pixel value $X_i$. The simulated micrographs are generated with some randomly initialized edges and later those edges are cut off so that only stationary pixels remain. Note that the two micrographs shown in Figures~\ref{fig_SM:ar_model1} and~\ref{fig_SM:ar_model2} are generated by two sets of coefficients and a nonlinear or linear $h(\cdot)$ function and they appear similar to some real stochastic microstructures of materials, e.g., the sandstone micrographs in~\cite{li2018transfer} and silica-filled rubber matrix micrographs in~\cite{bostanabad2016stochastic}. Denote by $\bm{\phi}^{\mathrm{(CL)}}$ and $\bm{\phi}^{\mathrm{(M)}}$ the $2$D AR coefficients for generating micrographs for the CL-selection data and the monitoring data, respectively. For the microstructures in Figures~\ref{fig_SM:2d_ar_sup_multi_chart_nnet} and~\ref{fig_SM:power_comp_nnet}, the parameters and configurations are: $c_0=1$; $ \sigma_{\mathrm{AR}}=0.01$; the row-by-row concatenated AR coefficient vectors $[\phi ^{\mathrm{(CL)}} _{0,0},\phi ^{\mathrm{(CL)}} _{0,1},\phi ^{\mathrm{(CL)}} _{0,2},\phi ^{\mathrm{(CL)}} _{1,0},\phi ^{\mathrm{(CL)}} _{1,1},\phi ^{\mathrm{(CL)}} _{1,2},\phi ^{\mathrm{(CL)}} _{2,0},\phi ^{\mathrm{(CL)}} _{2,1},\phi ^{\mathrm{(CL)}} _{2,2}]=[0,  3.59e-01, 1.07e-02, 3.90e-01, 4.21e-02, 1.76e-03, 9.98e-02, -1.82e-03, 1.72e-05]$ and $[\phi ^{\mathrm{(M)}} _{0,0},\phi ^{\mathrm{(M)}} _{0,1},\phi ^{\mathrm{(M)}} _{0,2},\phi ^{\mathrm{(M)}} _{1,0},\phi ^{\mathrm{(M)}} _{1,1},\phi ^{\mathrm{(M)}} _{1,2},\phi ^{\mathrm{(M)}} _{2,0},\phi ^{\mathrm{(M)}} _{2,1},\phi ^{\mathrm{(M)}} _{2,2}]=[0, 2.74e-1, 2.93e-2, -2.41e-1, 1.50e-1, -1.17e-2, 4.31e-1, 4.52e-2 ,-2.96e-2]$; and $h(x)=min(5, max(0.05, \exp (x)))$. For the microstructures in Figures~\ref{fig_SM:2d_ar_sup_multi_chart_nnet_1} and~\ref{fig_SM:power_comp_nnet_1}, the parameters and configurations are: $c_0=1$; $ \sigma_{\mathrm{AR}}=0.01$; the row-by-row concatenated AR coefficient vectors $[\phi ^{\mathrm{(CL)}} _{0,0},\phi ^{\mathrm{(CL)}} _{0,1},\phi ^{\mathrm{(CL)}} _{0,2},\phi ^{\mathrm{(CL)}} _{1,0},\phi ^{\mathrm{(CL)}} _{1,1},\phi ^{\mathrm{(CL)}} _{1,2},\\\phi ^{\mathrm{(CL)}} _{2,0},\phi ^{\mathrm{(CL)}} _{2,1},\phi ^{\mathrm{(CL)}} _{2,2}]=[0, 3.59e-01, 1.07e-01, 9.98e-03, -1.82e-03, 1.72e-05,  3.51e-01, 4.21e-02, 1.76e-03]$ and $[\phi ^{\mathrm{(M)}} _{0,0},\phi ^{\mathrm{(M)}} _{0,1},\phi ^{\mathrm{(M)}} _{0,2},\phi ^{\mathrm{(M)}} _{1,0},\phi ^{\mathrm{(M)}} _{1,1},\phi ^{\mathrm{(M)}} _{1,2},\phi ^{\mathrm{(M)}} _{2,0},\\\phi ^{\mathrm{(M)}} _{2,1},\phi ^{\mathrm{(M)}} _{2,2}]=[0, 3.59e-01, 1.07e-01, 9.98e-03, -1.82e-03, 1.72e-05, 3.12e-1, 4.21e-02, 1.76e-03]$; and $h(x)=x$.  

The control charts in Figure~\ref{fig_SM:2d_ar_sup_moni_multi_chart_nnet} show the results of the NM approach. We again fit a neural network model for $g(\bm{x}; \bm{\theta})$, using a single hidden-layer with $10$ nodes. Because here we can generate an arbitrary number of samples, we increase the number of CL-selection micrographs to $4$ with each micrograph of size $256\times 256$, which will allow us to more accurately control the false alarm rate when the CL-selection and monitoring micrographs are statistically equivalent (we only show a single CL-selection micrograph and a single monitoring micrograph in Figure~\ref{fig_SM:2d_ar_sup_moni_multi_chart_nnet} for brevity). We see that in terms of the power, the SWMA-$\bm { \theta}$ chart performs better than the SWMA-$\sigma$ chart in Figure~\ref{fig_SM:2d_ar_sup_multi_chart_nnet}, and the SWMA-$\sigma$ chart performs better than the SWMA-$\bm { \theta}$ chart in Figure~\ref{fig_SM:2d_ar_sup_multi_chart_nnet_1}. In both figures, the SWMA-M chart achieves the best performance of the SWMA-$\bm{\theta}$ chart and the SWMA-${\sigma}$ chart and is much better than the baseline RWMA chart.

To more comprehensively investigate both the power and the false alarm rate of our score-based approach, we gradually increase the difference between the CL-selection and monitoring micrographs. More specifically, we denote the two sets of $2$D AR coefficients used to generate any pair of CL-selection and monitoring micrographs in Figure~\ref{fig_SM:2d_ar_sup_moni_multi_chart_nnet} as $\bm{\phi}^{(p)} = \{\phi_{r,c}^{(p)}\}_{r,c=0}^{l_g}$ with $\phi_{0,0}^{(p)}=0$ and $p\in\{0,1\}$, where $p=0$ and $p=1$ denote reference and nonstationary micrographs, respectively. Then, we define a parameter $\gamma$, which governs how different the CL-selection micrographs are from the monitoring micrographs via:
\begin{align}
\begin{aligned}
\bm{\phi}^{\mathrm{(CL)}} =& \bm{\phi}^{(0)} \\
\bm{\phi}^{\mathrm{(M)}} =& (1-\gamma)\bm{\phi}^{(0)} + \gamma\bm{\phi}^{(1)}
\end{aligned}
\label{eqn_SM:coef_changing}
\end{align}
When $\gamma=0$, micrographs in the CL-selection and monitoring data are generated by the same $2$D AR model so that they are statistically equivalent, in which case the power is the false alarm rate $\alpha$. As $\gamma$ increases from $0$ to $1$, the images become increasingly statistically different. The results in Figure~\ref{fig_SM:2d_ar_sup_moni_multi_chart_nnet} correspond to $\gamma=1.0$.

In practice, if the sizes (in terms of total number of pixels) of the CL-selection data and the monitoring data are both sufficiently large, then the false alarm rate during monitoring will be the same as the desired value of $\alpha$ specified when selecting the control limits using the CL-selection data. This would give a common basis for comparison of the power of the different control charts. For all of our simulation experiments, we achieved a common false alarm rate differently, by bypassing the CL-selection data and selecting the control limits to directly control (at least approximately) the false alarm rate over a large set of monitoring data. From Figure~\ref{fig_SM:2d_ar_power_comp}, which plots the power of the different control charts as a function of $\gamma$, we can see that between the two component score-based charts, sometimes the SWMA-$\bm{\theta}$ chart has higher power than the SWMA-$\sigma$ chart, and sometimes vice-versa. And the SWMA-M chart always performs similarly to the best of the two component score-based control charts. In contrast, for this example the RWMA chart is completely ineffective at detecting the change and has a power that is not much higher than the false alarm rate even for the larger $\gamma$ values. Note that when $\gamma \cong 0.2$ in Figure~\ref{fig_SM:2d_ar_power_comp}, the difference of microstructures are difficult to discern with the human eye (we omit the micrographs for brevity), but our score-based method can still detect the differences with reasonable power, which further demonstrates its effectiveness. 

We also trained and used linear models for $g(\bm{x}; \bm{\theta})$ and found that the score-based method performed almost as well (in terms of power at detecting nonstationarity) as the neural network models, and so we omit the results for brevity. This is an interesting observation, because even though a linear model is not a correct model structure for $g(\bm{x};\bm{\theta})$ for this example (because of the nonlinear transformation $h(\cdot)$), our score-based framework still works reasonably well. Based on this observation, one potential strategy is to initially use a simpler model to take advantage of the lower computational expense of fitting the model and computing the score vectors, and then switch to a more complex model to get a better performance and interpretation if the simpler model signals a change. 

\subsection{Results for the ND Approach}
\label{ss_SM:res_unsupervised_chara}
We apply the ND approach to examples involving two different sets of micrograph data. The first data set consists of SEM images of dual-phase steel~(\cite{banerjee2013segmentation}), in which there are some martensite islands in ferrite matrix. Diagnosing/segmenting such multiphase (thus nonstationary) images is important for quality control and for understanding properties of steel samples. The second data set consists of TEM images of silica particles dispersed in PMMA with octyl functional modification. We choose images with different dispersion density and paste them together to form some artificial nonstationary microstructures and represent the practical problem of segmenting micrographs with multiple phases in each sample. Automatically segmenting regions with different microstructure characteristics like particle density is of interest to materials researchers, because the microstructure affects physical properties of the materials. We note that the ND approach can be applied to analyze and segment the different microstructures in a single multi-phase micrograph sample or in a collection of nonstationary micrograph samples.

\subsubsection{Dual-Phase Steel Data Set}
\label{sss_SM:res_score_dual_phase}

\begin{figure}[!htbp]
\captionsetup[subfigure]{width=.95\linewidth}
\centering
\includegraphics[width=.8\textwidth, trim=2in 2.3in 1.5in 1.5in, clip]{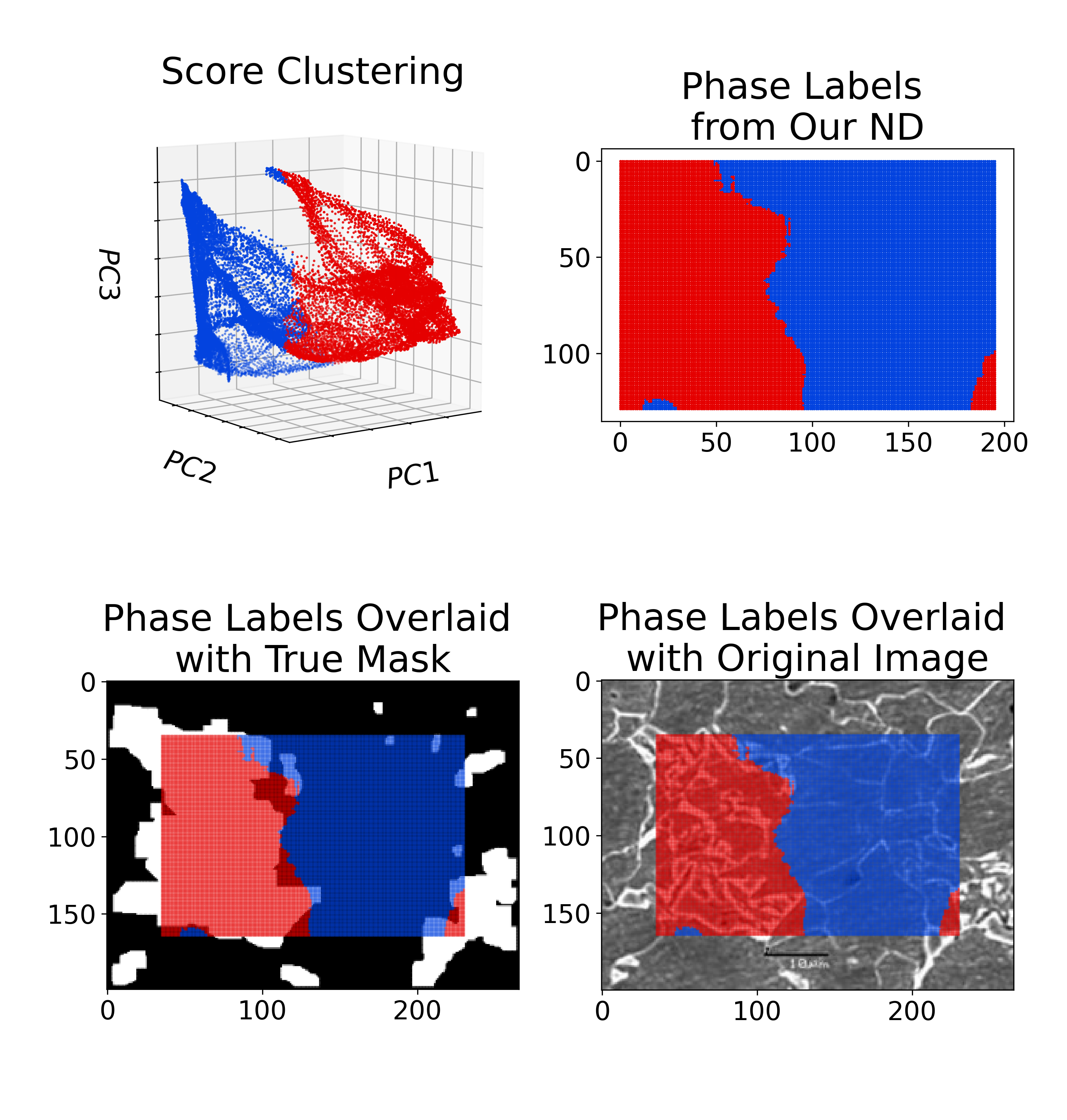}
  \caption{Results of our score-based method applied to an SEM image of a dual-phase steel sample. We used a neural network model with a single hidden-layer having $10$ neurons and parameters $l_s = 5$ and $l_w=30$. The color-coded predicted labels (top-right panel) are overlapped (bottom-left panel) with the mask showing the ``true" phases provided by~\cite{banerjee2013segmentation}. The bottom-right panel is the original SEM micrograph overlaid with the labels from our ND approach. The top-left plot are the first three PCA score for the set of score vectors, color-coded by ND phase labeling, which is useful for understanding the distribution of PCA components of score vectors in a high-dimensional space and debugging the data preprocessing and calculations of score vectors. The electronic version has color images of higher resolution.}
  \label{fig_SM:dual_phase_score_unsupervised}
\end{figure}

Figure~\ref{fig_SM:dual_phase_steel} shows an SEM image of dual-phase steel consisting of a ferrite matrix with martensite in the form of islands. This micrograph is especially challenging to analyze because some parts of the martensite regions are similar to parts of the ferrite regions and because the martensite areas are not connected. A human attempt to draw boundaries around the phase regions is very time-consuming, tedious, and error-prone. In contrast, the results of our score-based method shown in Figure~\ref{fig_SM:dual_phase_score_unsupervised} are very effective at identifying the multiple phases that are present and distinguishing them, in addition to being fully automated. In particular, the boundaries obtained from our score-based method are smooth and closely aligned with the true boundaries between the phases. In addition, the non-connected regions of martensite are successfully detected. The phase labels from our ND method are consistent with the mask, which can be treated as the ground truth, except that the boundaries of our labeled phase regions are smoothed to some extent. Similarly, some of the small islands of martensite have been smoothed out. The smoothing is the inevitable consequence of our WMA window and the neighborhood window having size larger than a single pixel. We emphasize that our score-based method is highly automated and uses minimal human intervention, while the method of~\cite{banerjee2013segmentation} requires many steps and heavy human-involved pre-processing, tuning, and post-processing.

\subsubsection{PMMA Data Set}
\label{sss_SM:res_score_silica_PMMA}
\begin{figure}[!htbp]
\captionsetup[subfigure]{width=.95\linewidth}
\centering
   \begin{subfigure}[t]{0.25\linewidth}
    \centering
	\includegraphics[width=\textwidth, trim=0.1in 0.1in 0.1in 0.1in, clip]{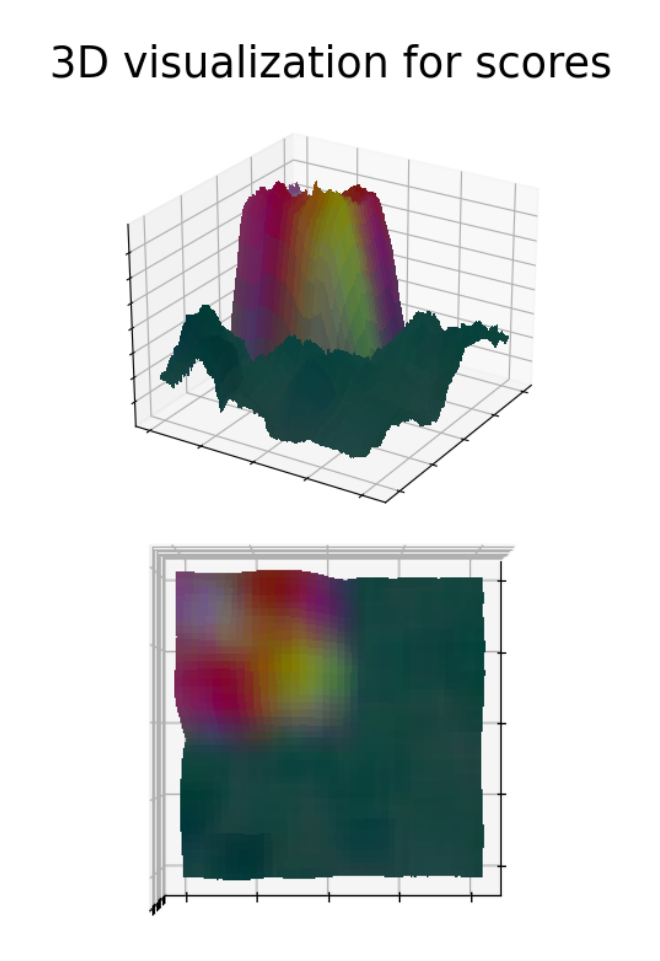}
         \caption{Visualization of the score vectors of a linear model from side (top panel) and top views (bottom panel).}
         \label{fig_SM:silica_PMMA_visu}
  \end{subfigure}
   \begin{subfigure}[t]{0.34\linewidth}
         \centering
         \includegraphics[width=\textwidth, trim=1.8in 1.8in 1.8in 0.in, clip]{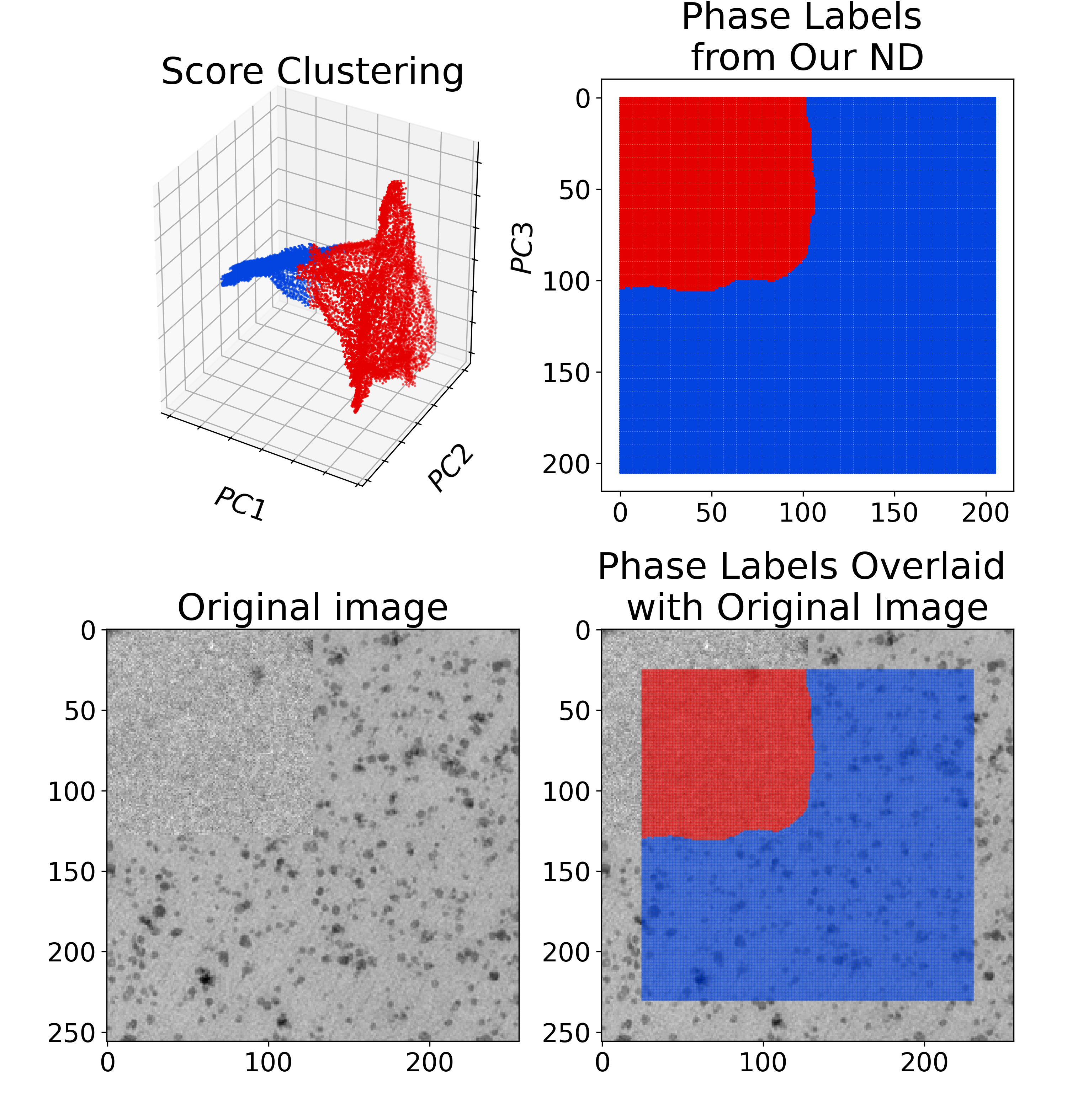}
         \caption{The results of the our score-based ND approach with a linear model.}
         \label{fig_SM:silica_PMMA_lin_unsupervised_seg}
  \end{subfigure}
  \begin{subfigure}[t]{0.34\linewidth}
         \centering
         \includegraphics[width=\textwidth, trim=1.8in 1.8in 1.8in 0.in, clip]{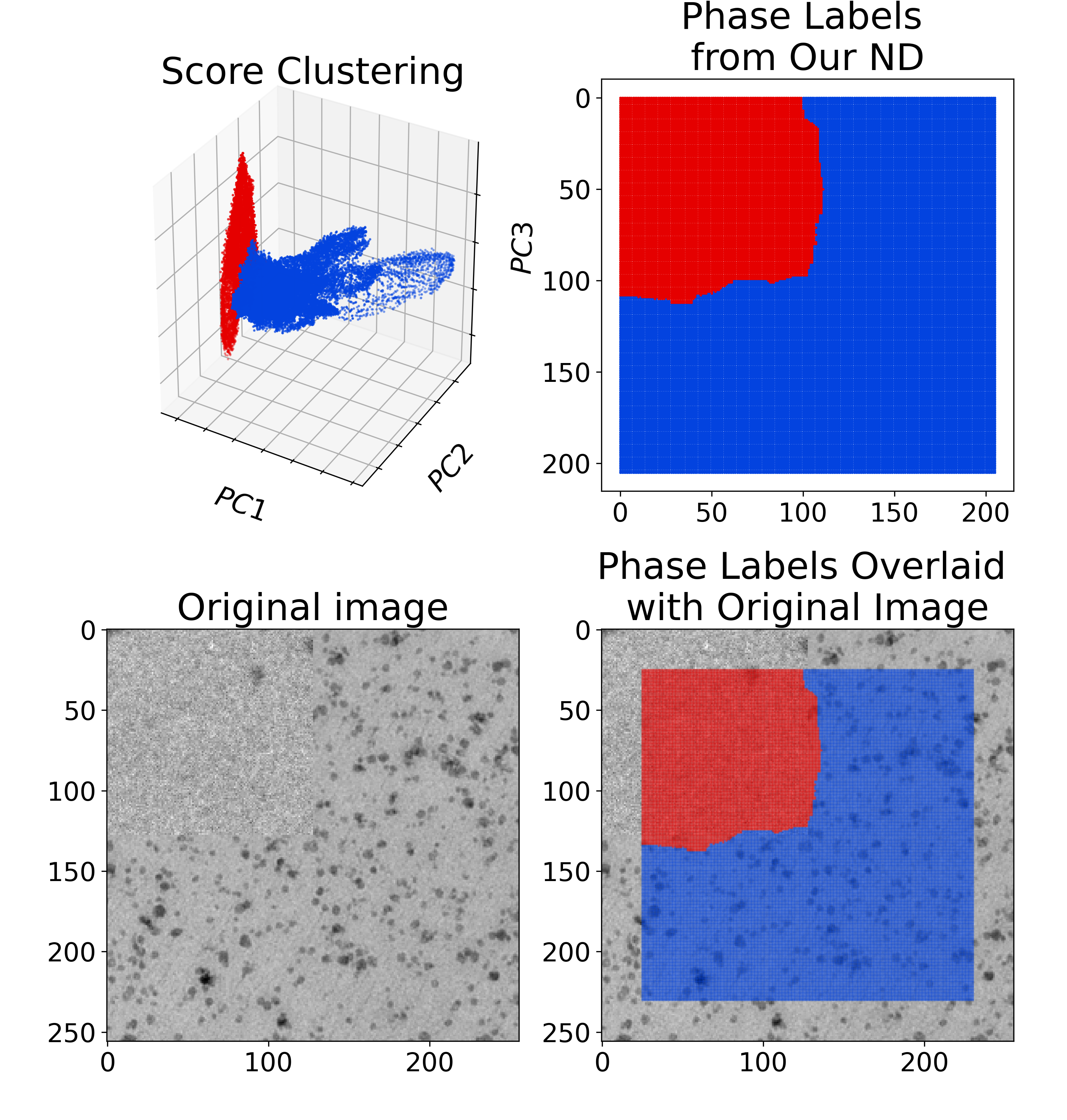}
         \caption{The results of the our score-based ND approach with a neural network model with one hidden-layer having $10$ nodes.}
         \label{fig_SM:silica_PMMA_nnet_unsupervised_seg}
  \end{subfigure}
  \caption{Results of our score-based method applied to a TEM image with nonstationarity generated by pasting together silica-PMMA materials samples with different particle dispersion densities. Here, we use parameters $l_s = 5$ and $l_w=20$. The electronic version has color images of higher resolution.}
  \label{fig_SM:silica_PMMA_score_unsupervised}
\end{figure}

The second data set consists of TEM images of silica particles dispersed in PMMA, some of which were shown in Figure~\ref{fig_SM:PMMA} and analyzed in Section~\ref{ss_SM:res_supervised_moni}. We created nonstationary micrographs by pasting a micrograph with low particle dispersion density to the upper left quadrant and three micrographs with higher particle dispersion density to the other three quadrants. As shown in Figure~\ref{fig_SM:silica_PMMA_visu}, the result of our visualization method with $3$D height and color mapping described in Section~\ref{s_SM:unsupervised_chara} implies there are mainly two kinds of microstructures in the sample. With this information, we applied our ND score clustering method with two clusters. Figures~\ref{fig_SM:silica_PMMA_lin_unsupervised_seg} and~\ref{fig_SM:silica_PMMA_nnet_unsupervised_seg} show that our score-based ND approach with either the linear model or the neural network as the supervised learner $g(\bm{x},\bm{\theta})$ can effectively diagnose the nonstationarity and accurately label the phases within the micrograph. The $3$D clustering figures also help visualize the distribution of PCA components of score vectors in a high-dimensional space to help understand how score vectors are different for different material phases and also help in debugging the data preprocessing and calculations of score vectors.

In practice, users can compare results using larger or smaller WMA windows, where a larger window will smooth out more noise and make the ND method better able to detect small microstructure differences that are sustained over larger spatial regions, but this will also tend to smooth out more localized nonstationarities. In contrast, using a smaller WMA window will be more sensitive to localized nonstationarities, but will be more sensitive to noise and less accurately detect small but sustained microstructure differences. In our investigation, we found that compared with the score-based ND approach, the score-based NM approach in Section~\ref{s_SM:supervised_moni} tends to indicate differences between different phases with higher power. This makes sense, because when the model is trained on a stationary micrograph(s) of a single phase, the variance of the score vectors will typically be smaller than when the model is trained on nonstationary micrograph(s) with multiple phases, and hence it becomes easier to detect differences between a new phase and the reference phase. On the other hand, the NM approach requires more data in the sense that one must have one or more reference micrographs available. Moreover, the ND approach provides more diagnostic information than the NM approach and explicitly identifies the micrograph regions corresponding to the multiple phases. 

\section{Conclusions}
\label{s_SM:conclusion}

In this study, we have developed a powerful and versatile score-based framework for nonstationarity analysis of stochastic microstructures of materials. This problem is of increasing importance due to the increasing availability of complex multiphase micrograph data and the lack of effectiveness of traditional methods. Modeling the stochastic nature through parametric supervised learning models and analyzing nonstationarity through our score-based framework have a solid theoretical foundation and, as we demonstrated with a number of examples, are effective and efficient at monitoring and diagnosing nonstationarity. We have developed two components in our framework: nonstationarity monitoring and nonstationarity diagnostics, which are intended for use in two common related but different practical scenarios. The consistently good performance across our examples and the higher statistical power that we demonstrated through Monte Carlo simulations are evidences of the effectiveness of our score-based approach and its advantages over the residual-based benchmark existing approach. The framework has substantial potential for automating and improving image analysis of materials microstructures and can be combined with other state-of-the-art machine learning and deep learning techniques, e.g., classification and segmentation deep learning networks, which we are currently investigating as extensions.  

\section*{Acknowledgement}
This work was funded in part by the Air Force Office of Scientific Research Grant \# FA9550-18-1-0381, which we gratefully acknowledge. The micrographs of silica particles in PMMA are courtesy of Prof. Linda Schadler (Linda.Schadler@uvm.edu) and Prof. Cate Brinson (cate.brinson@duke.edu). This work used the Extreme Science and Engineering Discovery Environment (XSEDE)~\cite{towns2014xsede}, which is supported by National Science Foundation grant number ACI-1548562, and the Quest high performance computing facility at Northwestern University which is jointly supported by the Office of the Provost, the Office for Research, and Northwestern University Information Technology.


\bibliography{mybibfile}

\end{document}